\documentclass[fleqn,usenatbib]{mnras}

\usepackage[T1]{fontenc}

\DeclareRobustCommand{\VAN}[3]{#2}
\let\VANthebibliography\thebibliography
\def\thebibliography{\DeclareRobustCommand{\VAN}[3]{##3}\VANthebibliography}

\usepackage{graphicx}	
\usepackage{amsmath}	
\usepackage{amssymb}	
\usepackage{xcolor}
\usepackage{lipsum}
\usepackage[english]{babel}
\usepackage{blindtext}
\usepackage{scalefnt}
\usepackage{soul}
\usepackage{multirow}
\usepackage{makecell}
\usepackage{arydshln}
\usepackage{float}
\usepackage{siunitx}
\usepackage{booktabs}
\usepackage{caption}

\newcommand{\LAGN}{$L_{\rm{AGN}}$}
\newcommand{\vAGN}{$v_{\rm{AGN}}$}

\newcommand{\kmin}{$k_{\rm{min}}$}
\newcommand{\Lmax}{$\lambda_{\rm{max}}$}
\newcommand{\Mtarget}{$M_{\rm{target}}$}

\newcommand{\LbolEq}[2]{\ensuremath{L_{\rm{AGN}}{#1} 10^{#2}\rm{\ erg\ s^{-1}}}}
\newcommand{\LAGNEq}[2]{\ensuremath{L_{\rm{AGN}}{#1} 10^{#2}\rm{\ erg\ s^{-1}}}}

\newcommand{\kminEq}[2]{\ensuremath{k_{\rm{min}}{#1}{#2}\rm{\ kpc^{-1}}}}
\newcommand{\LmaxEq}[2]{\ensuremath{\lambda_{\rm{max}}{#1}{#2}{\ \rm{pc}}}}
\newcommand{\MtargetEq}[2]{\ensuremath{M_{\rm{target}}{#1}{#2}{\ \rm{M_\odot}}}}
\newcommand{\vAGNEq}[2]{\ensuremath{v_{\rm{AGN}}{#1}{#2}{\rm{\ km\ s^{-1}}}}}
\newcommand{\vrEq}[2]{\ensuremath{v_{\rm{r}}{#1}{#2}{\rm{\ km\ s^{-1}}}}}
\newcommand{\vminEq}[2]{\ensuremath{v_{\rm{min}}{#1}{#2}{\rm{\ km\ s^{-1}}}}}
\newcommand{\TEq}[2]{\ensuremath{T{#1}10^{#2}{\rm{\ K}}}}
\newcommand{\tEq}[2]{\ensuremath{t{#1}{#2}{\rm{\ Myr}}}}

\newcommand{\pEq}[2]{\ensuremath{\dot{p}/(L/c){#1}{#2}}}

\definecolor{green}{RGB}{34, 139, 204}

\usepackage{newtxtext,newtxmath}

\usepackage{orcidlink}

\defcitealias{Ward2024}{Paper~I}

\title[X-ray emission from wind-ISM mixing]{Mixing between AGN winds and ISM clouds produces luminous X-ray emission}

\author[S.R. Ward et al.]{
S. R. Ward\,\textsuperscript{\orcidlink{0000-0001-5345-0900}}$^{1,2}$\thanks{E-mail: \url{sward@flatironinstitute.org}}, 
T. Costa\,\textsuperscript{\orcidlink{0000-0002-6748-2900}}$^{2}$, 
C. M. Harrison\,\textsuperscript{\orcidlink{0000-0001-8618-4223}}$^{2}$, and
V. Mainieri\,\textsuperscript{\orcidlink{0000-0002-1047-9583}}$^{3}$
\\
$^{1}$Center for Computational Astrophysics, Flatiron Institute, 162 5th Avenue, New York, NY 10010, USA \\
$^{2}$School of Mathematics, Statistics and Physics, Newcastle University, Newcastle upon Tyne, NE1 7RU, UK \\
$^{3}$European Southern Observatory, Karl-Schwarzschild-Straße 2, 85748 Garching bei München, Germany\\
}

\date{Accepted to MNRAS, 2025-11-18}

\pubyear{2025}

\begin{document}
\label{firstpage}
\pagerange{\pageref{firstpage}--\pageref{lastpage}}
\maketitle

\begin{abstract}

Active galactic nuclei (AGN) drive powerful, multiphase outflows that are thought to play a key role in galaxy evolution.
The hot, shocked phase of these outflows (\TEq{\gtrsim}{6}) is expected to dominate the energy content, but is challenging to observe due to its long cooling time and low emissivity. The cool phase (\TEq{\lesssim}{4}) is easier to detect observationally, but it traces a less energetic outflow component.
In prior simulations of the interaction between an energy-driven AGN outflow and a clumpy ISM, we found that mixing between hot wind and cool ISM clouds produces a new, highly radiative, phase at \TEq{\approx}{6-7} which fuels the formation of a long-lived ($\geq 5\ \rm{Myr}$) cool outflow. We investigate the X-ray emission generated by thermal Bremsstrahlung and high-ionisation metal line emission in this mixing phase, finding that it could contribute significantly to the X-ray output of the outflow. This mixing-induced X-ray emission is strongest in the part of the outflow propagating equatorially through the disc, and is extended on scales of $D\simeq 3-4\ \rm{kpc}$. For quasar luminosities of \LAGNEq{\simeq}{45-46}, the resulting X-ray luminosity is equivalent to that expected from star formation rates $\rm{SFR}\simeq 10-200\ \rm{M_\odot\ yr^{-1}}$, showing that it could be an important source of soft X-rays in AGN host galaxies.
Our results suggest that this extended emission could be resolvable in local quasars ($z\lesssim 0.11$) using high spatial-resolution X-ray observatories such as \textit{Chandra}, or proposed missions such as \textit{AXIS} and \textit{Lynx}.



\end{abstract}




\section{Introduction}
\label{sec:intro}

Feedback from active galactic nuclei (AGN) has become an essential component of modern cosmological models and simulations of galaxy evolution \citep[e.g.,][]{Springel2005_BH,Somerville2008,Schaye2015,Dubois2016,Weinberger2018,Dave2019,Dubois2021,Wellons2023,Dolag2025}, with the energy released from accreting supermassive black holes (SMBHs) thought to play a key role in quenching star formation in massive galaxies \citep{Scannapieco2004,Churazov2005,Bower2006,Fabian2012,Terrazas2017,Piotrowska2022,Ward2022}. However, a key uncertainty in models of feedback is how efficiently the energy from the AGN can couple to the galaxy's multiphase interstellar medium (ISM). Understanding this coupling efficiency is therefore key to uncovering the impact of AGN on their host galaxies, but constraining this quantity from observations is highly challenging \citep[see discussions in][]{Harrison2018,Veilleux2020,Harrison2024,Ward2024}.

To assess the impact of AGN, studies target kiloparsec-scale outflows, which are launched when accretion disc winds, jets and/or radiation pressure from the AGN sweep up the surrounding ISM and accelerate it to high velocities (\vrEq{\gtrsim}{100}). These outflows are multiphase, containing entrained clouds of molecular or neutral gas \citep[e.g.,][]{Gonzalez-Alfonso2017,Fluetsch2019,Veilleux2020,Lamperti2022,RamosAlmeida2022}, warm ionised gas \citep[e.g.,][]{Rose2018,ForsterSchreiber2019,Polack2024,Bertola2025}, and hot X-ray emitting gas \citep[e.g.,][]{Tombesi2013,Greene2014,Lansbury2018}. Therefore, to fully characterise the outflow, and thus infer its effect on the galaxy, a multiwavelength approach is needed to capture all the mass and energy \citep{Cicone2018,Girdhar2022,Speranza2024,Harrison2024}. In particular, the hot ionised phase (\TEq{\gtrsim}{6}) is expected to dominate the energy budget of the outflow, although the long cooling time of this phase means it is not expected to radiate efficiently.

A strong candidate for creating the kiloparsec-scale outflows seen in observations (including in X-rays) are accretion disc winds. In this picture, the accretion disc launches a small-scale wind which interacts with the surrounding medium, shocking it to high temperatures, and creating a large-scale outflow. An analytic model for this process was introduced in \cite{King2003,King2005} and developed further in \cite{Zubovas2012,Faucher-Giguere2012} and \cite{Costa2014b}. 
\cite{Nims2015} provided analytic arguments for observational signatures of this model, discussing both thermal (inverse-Compton and Bremsstrahlung) and non-thermal (synchrotron, non-thermal inverse-Compton and pion decay) mechanisms. 
A numerical implementation in the hydrodynamic code \textsc{Arepo} was presented in (\textsc{Bola}; \citealt{Costa2020}). 

In our previous work \citep{Ward2024} we placed the \textsc{Bola} AGN model at the centre of a gas disc with a fractally-distributed clumpy ISM structure and studied the resulting multiphase outflow. We found that the resulting outflow differed significantly from previous analytic work that had assumed a smooth background. The outflow consisted of a high-velocity hot phase \TEq{\gtrsim}{6} that could stream freely out of low density regions and which contained most of the energy of the outflow, and slower moving (\vrEq{\lesssim}{400}) cold (\TEq{\leq}{4.5}) clouds that could survive in the outflow by mixing with the hot wind, boosting their radiative cooling. These results motivated us to investigate the predicted X-ray emission from our simulations.

We are focussing this paper on X-ray emission for two main reasons: the first is that, as we found in \cite{Ward2024}, the hot phase of the outflow carries the bulk of the energy. This makes it critical to understand but, its low density and long cooling time make it hard to observe. However, as it mixes with the colder ISM phases, it cools radiatively, allowing us to observe the interaction and indirectly study this otherwise invisible gas phase. Additionally, we found that the survival of cold gas in our galactic outflows was likely dependent on this mixing between the hot and cold phases, which boosts the radiative cooling. This effect has also been seen in wind-tunnel simulations \citep[see e.g.][]{Gronke2020,Fielding2020}. This mixing processes produces X-ray emission via thermal Bremsstrahlung and high-ionisation metal lines due to the typical temperature of this mixed gas (\TEq{\simeq}{6-7}).

There have been several observational successes in detecting kiloparsec-scale hot gas in AGN outflows using X-ray imaging, mostly from the \textit{Chandra} and \textit{XMM-Newton} space telescopes (e.g., \citealt[][]{Croston2008,Feruglio2013,Greene2014,DiGesu2017,Lansbury2018,Fabbiano2022,TrindadeFalcao2025} and see review in \citealt{Fabbiano2022chapter}). However it is not clear whether the emission seen in these systems is caused by cooling from a shock-heated outflow bubble, photoionisation caused by an AGN, or star formation \citep[e.g.,][]{Wang2010,Somalwar2020,Wang2024}. The Milky Way also contains kiloparsec-scale X-ray bubbles \citep{Predehl2020} as well as narrow `chimneys' linking them to the galactic centre \citep{Ponti2019,Ponti2021}, making even our own galaxy a potential test-bed for signatures of AGN feedback. However, there are only limited theoretical or numerical predictions for the expected observational signatures of outflowing X-ray emitting gas and the relative contribution to the total X-ray emission compared to other processes such as star formation. 

Previous theoretical work has predicted that outflows should produce observable X-ray emission; for example, in their analytic work on quasar winds, \cite{Nims2015} concluded that the shocked ambient medium should produce X-ray emission on scales of a few kiloparsecs via Bremsstrahlung radiation. Additionally, there have been some numerical works that have predicted extended X-ray emission in the circumgalactic medium (CGM) of galaxies  \citep{Costa2014a,Pillepich2021,Bennett2024}. However, as shown in \cite{Ward2024}, outflow properties can differ significantly from a homogeneous case when the parsec-scale structure of the ISM is taken into account. Thus it is unclear what the effect of a multiphase, inhomogeneous ISM structure would be on the resulting X-ray emission.

In this study, we investigate the X-ray emission from interactions between an AGN wind and a clumpy ISM, building on our previous work \citep{Ward2024} where we introduced a series of numerical experiments featuring an AGN wind model embedded in both a smooth and clumpy ISM. This paper is structured as follows: in Section \ref{sec:methods} we summarise the simulations used in this study and describe how we estimate the X-ray emission; in Section \ref{sec:res} we show the main results of our work, including the effect of introducing metal line cooling in our simulations; in Section \ref{sec:discussion} we discuss whether this X-ray emission could be observable in local quasars; and finally in Section \ref{sec:conc} we summarise our findings.

We assume a flat, $\Lambda$CDM cosmology throughout, using values from \cite{Planck2016cite} of $H_0 = 67.7 \rm{\ km s^{-1}\ Mpc^{-1}}$, $\Omega_m = 0.3$ and $\Omega_\Lambda = 0.7$.

\section{Methods} \label{sec:methods}

In this section, we briefly describe the setup of the simulation suite and discuss how we estimate the X-ray emission from the outflow. For full details of the simulations, the reader is referred to \cite{Ward2024}, hereafter \citetalias{Ward2024}.

\subsection{The ACDC simulations}

In this study, we use the ACDC (AGN in Clumpy DisCs) simulation suite presented in \citetalias{Ward2024}. These simulations model an isotropic AGN wind situated within a galactic disc with a manually-distributed clumpy ISM. This idealised setup allowed us to perform a series of controlled experiments investigating how disc sub-structure affects the propagation of multiphase outflows. While the present paper focuses on the hot and mixed phases of the outflow, Almeida et al. 2025 (sub. MNRAS) examine the distribution and densities of the cold clouds entrained within the flow.

The simulations are performed using the moving-mesh hydrodynamic code \textsc{Arepo} \citep{Springel2010,Pakmor2016,Weinberger2020} which features an unstructured Voronoi mesh that moves with the fluid and refines/de-refines in regions of high/low density. An exact Riemann solver is used at the cell interfaces to calculate hydrodynamic fluxes between each cell. This yields both accurate shock-capturing and high spatial resolution in dense regions.

\subsubsection{Clumpy disc setup}

It is computationally challenging to produce a realistic ISM structure in galaxy simulations. Therefore we follow studies such as \cite{Sutherland2007,Cooper2008,Wagner2011,Mukherjee2016,Bieri2017,BandaBarragan2020,Tanner2022} in manually setting the spatial distribution and phase structure of the ISM. This is achieved using the PyFC\footnote{\url{https://pypi.org/project/pyFC/}} package \citep{Wagner2012} to create a log-normal density distribution with a fractal spatial distribution \citep{LewisAustin2002,Sutherland2007}. This creates a random distribution of cold, dense clouds, with the clump size parameterised by the average largest cloud size, \Lmax. We investigate three such cloud sizes: \LmaxEq{=}{40} (`small' clouds); \LmaxEq{=}{170} (`medium' clouds); and \LmaxEq{=}{330} (`large' clouds). We also show the results for a `smooth' disc which has a homogeneous density distribution for comparison.

We use a disc with diameter $4\ \rm{kpc}$ and height $1\ \rm{kpc}$. The initial mean gas number density of the disc is set to $\langle n_0\rangle=5\ \rm{cm^{-3}}$ and the mean temperature to $\langle T_{0,\rm{disc}}\rangle= 10^4\ \rm{K}$, leading to initial clumps with densities $n_0 = 10^{1-3}\ \rm{cm^{-3}}$. An initial step is performed where cells with temperatures above $T_{\rm{crit}}=3 \times 10^4\ \rm{K}$ are replaced by hot background gas, to generate porosity in the disc. The background is a static halo in pressure equilibrium with the disc and an initial constant temperature and density of $T_{0,\rm{bkg}}= 10^7\ \rm{K}$ and $n_{0,\rm{bkg}}= 10^{-2}\ \rm{cm^{-3}}$ respectively.

\subsubsection{Cooling models} \label{sec:line cooling}

The original simulations in \citetalias{Ward2024} were performed assuming primordial cooling (including, e.g., Bremsstrahlung and radiative recombination.) However, in this study, we perform new simulations which include metal-line cooling. This was partially motivated by the need to include emission from metal lines when calculating the X-ray spectra (see \ref{sec:method spectra}).
Metal lines will not only affect the predicted X-ray emission from the simulations, but also provide an additional cooling channel that may impact the phase and spatial structure of the outflow. For this reason, we re-simulated a subset of our simulations with metal transport and metal-line cooling included. For these simulations, we altered our initial conditions to include a metallicity field. We initialised our clumps with a metallicity of $Z/Z_\odot=1$ and set the hot background to $Z/Z_\odot=0.1$, assuming solar abundances. The simulations were then run in the same way as the pre-existing models, but with metal-line cooling activated \citep{Vogelsberger2013}. We discuss the impact of metal-line cooling on our outflow structure in Section \ref{sec:spectra} and discuss how it impacts the global outflow properties investigated in \citetalias{Ward2024} in Appendix \ref{sec:ap:Zcooling}.

\begin{table*}
\begin{tabular}{@{}lrrrrrrrrr@{}}
\toprule
Simulation Name                  & {\Mtarget} & {\kmin}  & {\Lmax} & $h$   & $\langle n_0 \rangle$  & {\LAGN} & {\vAGN} & Cooling models \\ 
 &                  [$M_\odot$] & [$\rm{kpc}^{-1}$]  & [$\rm{pc}$] & [$\rm{kpc}$]    & [$\rm{cm^{-3}}$]    & [$\rm{erg\ s^{-1}}$]  & [$\rm{km\ s^{-1}}$] &  \\ \midrule

Small clumps &  100             & 50               & 40              & 1          & 5                                        & $10^{45}$         & 10000     & Primordial, metal line*     \\
\textbf{Medium clumps (fiducial)} &  \textbf{100}    & \textbf{12}      & \textbf{170}    & \textbf{1} & \textbf{5}         & $\mathbf{10^{45}}$ & \textbf{10000} & \textbf{Primordial, metal line*} \\
Large clumps &  100             & 6                & 330             & 1          & 5                                        & $10^{45}$         & 10000         &  Primordial \\
Smooth &  100             & - & -               & 1          & 5                                        & $10^{46}$         & 10000       & Primordial   \\
 Slow wind&  100             & 12               & 170             & 1          & 5                                        & $10^{45}$          & 5000         & Primordial  \\
  Fast wind*&  100             & 12               & 170             & 1          & 5                                        & $10^{45}$          & 30000         & Primordial, metal line  \\
 Thin disc &  100             & 12               & 170             & 0.5        & 5                                        & $10^{45}$          & 10000       & Primordial   \\
 Low density &  100             & 12               & 170             & 1          & 2.5                                         & $10^{45}$          & 10000      & Primordial    \\
  High density* &  100             & 12               & 170             & 1          & 50                                         & $10^{45}$          & 10000      &  Primordial    \\
 L43 &            100             & 12               & 170             & 1          & 5                                        & $10^{43}$         & 10000        & Primordial, metal line*  \\
 L44&  100             & 12               & 170             & 1          & 5                                        & $10^{44}$         & 10000       & Primordial, metal line*   \\
 L46&  100             & 12               & 170             & 1          & 5                                        & $10^{46}$        & 10000      & Primordial, metal line*    \\
 L47&  100             & 12               & 170             & 1          & 5                                        & $10^{47}$         & 10000     & Primordial, metal line*     \\
 No AGN &  100             & 12               & 170             & 1          & 5                                        & -                               & -         & Primordial, metal line*     \\
Low-res &  1000            & 12               & 170             & 1          & 5                                        & $10^{45}$          & 10000      & Primordial    \\
 High-res &  10            & 12               & 170             & 1          & 5                                       & $10^{45}$          & 10000       & Primordial   \\ \bottomrule
\end{tabular}
\caption{The ACDC suite of simulations explored in this work. Asterisks mark new simulations since {\protect \cite{Ward2024}}. The columns represent: 1) The nickname for each run; 2) The target mass resolution; 3) The wavenumber of the initial clump sizes; 4) The average maximum cloud size of the clumps; 5) The height of the disc; 6) The initial mean number density of the clumps; 7) The luminosity of the AGN; 8) The velocity of the AGN wind at injection; and 9) The cooling models used.}
\label{table:runs}
\end{table*}

\subsubsection{AGN wind model}

We use the \textsc{Bola} wind model \citep{Costa2020} to generate a fast, small-scale wind. Two spherical shells of cells are fixed in place at the centre of the disc and mass, momentum and energy are then injected across the boundary of these cells at a scale of 10 pc. We model a spherical, ultra-fast outflow (UFO), with a momentum boost factor of $\tau=\dot{p}/(L_{\rm{AGN}}/c)=1$ and fiducial values for the wind velocity of $v_{\rm{AGN}}=10^4\ \rm{km\ s^{-1}}$, and an AGN luminosity of \LAGNEq{=}{45} (we investigate the effect of luminosity on the resulting X-ray emission in Sections \ref{sec:time_evo} \& \ref{sec:scaling}). The initial kinetic luminosity of the wind is given by:

\begin{equation}
    \dot{E}_w = \frac{\tau }{2c} v_{\rm{AGN}} L_{\rm{AGN}}
\end{equation}

which for our fiducial AGN values gives an injection-scale energy coupling efficiency of $\dot{E}_w/L_{\rm{AGN}} = 1.7 \%$. These values are consistent with observational studies of UFOs \citep[e.g.,][]{Gofford2015,Matzeu2023} and have been shown to lead to energy-driven outflows \citep[e.g.,][]{King2003,Faucher-Giguere2012,Costa2014b,Costa2020,Ward2024}. Although the wind is spherical at the injection scale, the shape of the disc collimates it into a bipolar outflow \citep[see][]{Costa2020}.

A passive scalar is also injected along with the wind ($\mathcal{P}$) in order to trace this component throughout time. This parameter represents the mass fraction of injected wind material in the cell. An additional refinement scheme is imposed to further refine the cells in regions of high wind tracer density to increase the resolution in the wind by a factor of 10.

\subsubsection{Simulation suite}

The simulations used in this study are summarised in Table \ref{table:runs}. We show the target mass resolution ($M_{\rm{target}}$, where a smaller mass resolution results in finer spatial resolution), as well as various disc properties, such as the initial sizes of the fractal clumps (parameterised by average maximum size of a cloud; $\lambda_{\rm{max}} = 2 / k_{\rm{min}}$), the disc height ($h$), and the mean gas number density ($\langle n_{\rm{0}} \rangle$). We also show the properties of the injected AGN wind, namely the AGN luminosity ($L_{\rm{AGN}}$) and initial wind velocity ($v_{\rm{AGN}}$). In the final column, we show which cooling models the setups have been run with, whether just primordial, or also including a simulation with metal-line cooling.

New simulations since \citetalias{Ward2024} are marked with an asterisk; this includes all simulations run with metal line cooling. Our fiducial simulations are in bold, representing a medium clump size ($\lambda_{\rm{max}}=170\ \rm{pc}$) and an AGN luminosity and wind velocity of \LAGNEq{=}{45} and \vAGNEq{=}{10000} respectively. The effect of all these parameter variations on the total resulting X-ray emission is shown in Appendix \ref{sec:ap:params}.

Our fiducial mass resolution is \MtargetEq{=}{100} which gives us a spatial resolution down to $d_{\rm{cell}}\approx 1\ \rm{pc}$ in the highest density regions. In \citetalias{Ward2024}, we performed a numerical convergence test and found that global outflow properties are well-converged between our fiducial and high resolution (\MtargetEq{=}{10}) simulations. The effect of resolution on the results in this paper is discussed in Appendix \ref{sec:ap:res}.

\subsection{Calculating X-ray emission}

To estimate the X-ray emission in our simulations, we use two methods: the first is to use a simple approximation and assume only Bremsstrahlung emission, and the second involves computing the full X-ray spectra to include the effect of metal line emission.

\subsubsection{Bremsstrahlung emission}

The hot ionised gas phase (\TEq{\gtrsim}{6}) is expected to emit X-rays via thermal processes. The most important of these is Bremsstrahlung (free-free) radiation, emitted by free electrons interacting with charged ions. To estimate the bolometric X-ray luminosity from the hot gas, we use the Bremsstrahlung approximation \citep[see also][]{Sijacki2006,Costa2014a,Bennett2022}. The Bremsstrahlung emissivity is given by:


\begin{align}
    \epsilon_{\nu} &=  \frac{2^5 \pi e^6}{3 m_e c^3} \left( \frac{2 \pi}{3 k_B m_e} \right)^{\frac{1}{2}}    n_e n_i T^{-\frac{1}{2}} Z^2 \bar{g} e^{-\frac{h\nu}{k_B T}} \\
    &= 6.8\times 10^{-38}   n_e n_i T^{-\frac{1}{2}} Z^2 \bar{g} e^{-\frac{h\nu}{k_B T}} \quad \rm{[erg\ s^{-1}\ cm^{-3}\ Hz^{-1}]}
\end{align}

where $T$ is the gas temperature, $n_e$ and $n_i$ are the electron and ion number density respectively, $Z$ is the mean charge of the ions, $\nu$ is the frequency of the emitted radiation, and $\bar{g}\approx 1$ is the Gaunt factor. The luminosity per volume is then given by integrating over a frequency range:

\begin{align}
    \frac{dL_X}{dV} &= \int^{\nu_2}_{\nu_1} \epsilon_{\nu}  d \nu \\
    &= 1.4\times 10^{-27} n_e n_i  T^{\frac{1}{2}} Z^2 \bar{g} \left[ e^{-\frac{h\nu}{k_B T}} \right]^{\nu_2}_{\nu_1} \ \rm{[erg\ s^{-1}\ cm^{-3}]} \label{eq:freq int}
\end{align}

which for the bolometric case, $\nu \in [0,\infty )$, becomes

\begin{equation}
    \frac{dL_X}{dV} = 1.4\times 10^{-27}\ \ n_e\ n_i\ T^{\frac{1}{2}}\ Z^2\ \bar{g} \qquad \qquad \rm{[erg\ s^{-1}\ cm^{-3}]}
\end{equation}

The total X-ray emission is then evaluated as a sum over all the cells,

\begin{equation}
    L_X= 1.4\times 10^{-27}\ \bar{g} \sum  n_e n_i  T^{\frac{1}{2}} Z^2   V \qquad \qquad \rm{[erg\ s^{-1}]}
    \label{eq:brem}
\end{equation}

Alternatively, the frequency interval (Equation \ref{eq:freq int}) can be calculated over the standard soft ($\nu \in [0.5,2 ] \rm{\ keV}$) or hard X-ray bands ($\nu \in [2,10] \rm{\ keV}$).

This approximation assumes that Bremsstrahlung emission is the dominant source of radiation from the gas, neglecting thermal inverse-Compton scattering, radiative recombination and metal lines. \cite{Nims2015} calculated the expected emission from an analytic AGN wind and found that Bremsstrahlung emission from the outflow dominated inverse-Compton emission at the spatial scales ($\gtrsim 10 \ \rm{pc}$) and densities ($n \approx 10\ \rm{cm^{-3}}$) relevant in this work.  For partially-ionised gas with \TEq{\lesssim}{6}, radiative recombination (free-bound) becomes an important continuum mechanism in the soft band. These effects suggest that a Bremsstrahlung-only approximation of the X-ray emission represents a lower estimate, especially for softer ($\lesssim1\rm{\ keV}$) X-rays.

Nevertheless, estimating the emission using the Bremsstrahlung approximation is a useful tool to study the processes causing X-rays in the outflow as it is computationally straightforward, has minimal model dependencies, and allows comparison to previous analytic work \citep[e.g.,][]{Nims2015}. However, when comparing to observations, we further need to include contributions from metal lines and radiative recombination which can significantly boost the observed flux, especially in the soft band.

\begin{figure*}
    \centering
    \includegraphics[width=0.95\textwidth]{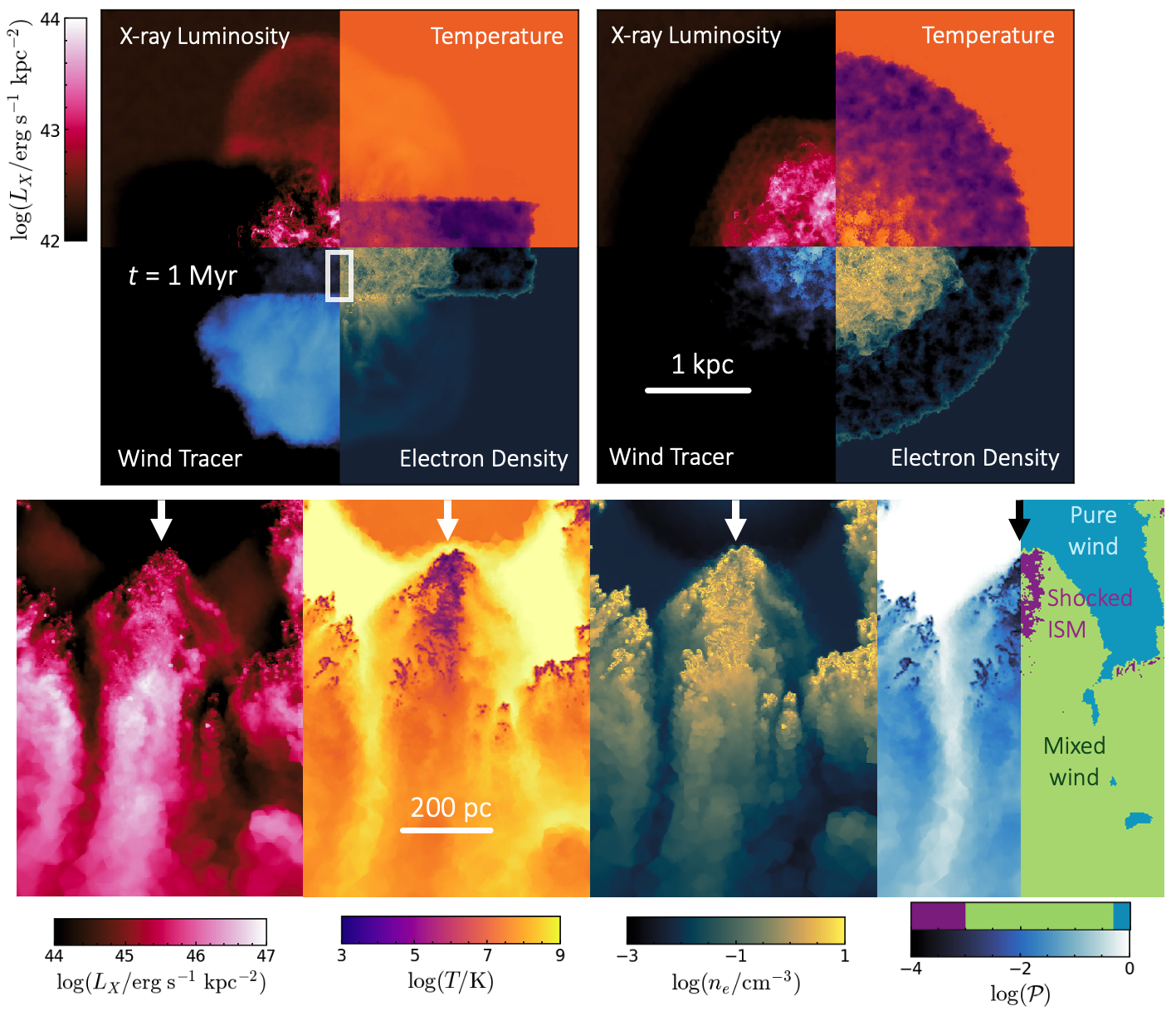}
    \caption{An overview of our simulations, showing the resulting X-ray luminosity from Bremsstrahlung emission at \tEq{=}{1}. The top row shows a $(5\ \rm{kpc})^3$ projection of the galaxy disc side-on (left) and top-down (right). Across all the panels, the X-ray luminosity has been integrated along the line of sight and the other quantities are density-weighted averages (temperature, electron density and wind tracer density). The bottom panels show a zoom-in of an initially dense gas cloud and the tail formed from stripped and mixed gas behind it. We find that the strongest X-ray emission due to the quasar-driven outflow comes from gas mixing with the wind in the tail behind the dense clump, forming a luminous chimney, on a scale of a few hundred parsecs.}
    \label{fig:overview}
\end{figure*}

\subsubsection{X-ray spectra} \label{sec:method spectra}

To go beyond the Bremsstrahlung approximation and calculate full X-ray spectra, we used the \textsc{PyAtomDB}\footnote{\url{atomdb.readthedocs.io}} package. This consists of a large atomic database for creating UV and X-ray spectra from collisionally ionised gas. We pre-computed a grid of spectra in $\log{T}-Z$ space, with bins of $\Delta \log{T}=0.2$ (for \TEq{\geq}{4.5}) and $\Delta Z=0.1$, assuming collisional ionisation equilibrium. Each simulation cell was then mapped to the nearest point on the $\log{T}-Z$ grid, giving us an estimated X-ray spectra for each cell which could then be integrated over the simulation domain for a global spectrum of the galaxy. We show the resulting spectra in Section \ref{sec:spectra} where we compare it to the Bremsstrahlung approximation described above.

\section{Results} \label{sec:res}

In this section we present the results of our X-ray analysis. We start by considering the Bremsstrahlung approximation as a simple tracer of the X-ray emitting gas. This has the advantage of having minimal model dependencies and allows us to compare to previous analytic work \citep{Nims2015} which only considered primordial cooling. In Section \ref{sec:spectra} we then investigate the impact of including metal line cooling on the outflow and resulting X-ray spectra.

\subsection{Wind-cloud interactions produce X-ray emission} \label{sec:sum_fig}

Figure \ref{fig:overview} shows the resulting X-ray emission from the fiducial simulations presented in \citetalias{Ward2024} at \tEq{=}{1}. The top row shows the full galaxy disc edge-on (left panel) and top-down (right panel). These panels show a projection of the integrated X-ray luminosity from Bremsstrahlung emission and a density-weighted mean of the temperature, electron density and wind tracer density, clockwise from top left. We can see that by \tEq{=}{1} the outflow has travelled about 1 kpc within the disc (equatorial outflow) and  around 2 kpc outside the disc into the halo (polar outflow). The equatorial outflow has the effect of heating (to \TEq{\approx}{6-7}) and compressing the gas within the disc (top-right subpanel), which increases the free electron density (bottom-right subpanel). This results in bright X-ray emission within the disc. As the gas distribution in the initial disc was clumpy, the X-ray emission is also inhomogeneous; we discuss how the initial clumps lead to bright X-rays in the next section. The polar outflow results in roughly symmetric X-ray bubbles rising in the halo. However, these are less bright than the X-ray emission from the equatorial outflow. We can see that the electron density of these bubbles is only slightly higher than the background, resulting in weak X-ray emission. However, this may be due to our static, isobaric background which is rather a simplistic model of the environment around galaxies. Therefore, in a more realistic galaxy halo, the outflow may be able to sweep up more gas in these polar bubbles, resulting in higher electron density and/or higher temperatures and thus higher X-ray luminosity. Exploration of this is left to future work.

Linking the disc to these halos, we see bright `chimneys' of X-ray emitting gas. In the lower panels of Figure \ref{fig:overview}, we show a closer view of one of these structures which represent some of the most X-ray luminous gas. We show slices of a $d \approx 100 \ \rm{pc}$ cloud and its tail, showing from left to right the integrated X-ray luminosity, and density-weighted averages of the temperature, electron density and wind tracer density. The AGN wind is approaching from the top. This cold cloud (\TEq{\approx}{4}) has survived for \tEq{\geq}{1} despite strong ram pressure from the wind. Behind the cloud, we can see a long tail of gas which is confined by the hot wind venting through either side of it at high velocity (\vrEq{\gtrsim}{2000}, see \citetalias{Ward2024}). However, despite these vents containing the hottest gas, their low density ($n_e\approx 10^{-2}\ \rm{cm^{-3}}$) results in low X-ray emission. We find that the X-ray emission is strongest in the region behind the cold clump (the `tail'), which has both a moderate density ($n_e\approx 1\ \rm{cm^{-3}}$) and temperature (\TEq{\approx}{6-7}). 

In the rightmost panel, we show the wind tracer density. This allows us to see the contribution of the injected wind to the X-ray luminosity. We split the wind tracer into three regimes: `pure wind' ($\mathcal{P}>0.5$) which is dominated by the injected AGN wind; `mixed wind' ($10^{-3}<\mathcal{P}<0.5$) where the injected wind has thoroughly mixed with the initial medium; and `shocked ISM' ($\mathcal{P}<10^{-3}$) which are regions which have only very weakly mixed with the wind fluid. We can see that the initial clump has low wind tracer values as the wind has not penetrated the dense gas and the region above the clump is dominated by the freely-expanding wind. We can also see the `vents' either side of the gas tail where the wind is escaping along low-density channels. This region has low X-ray luminosity, suggesting that strong X-ray emission anti-correlates with the locations where the AGN wind is venting through low-density regions. The region behind the cloud, where the brightest X-ray emission is produced, is in the mixed wind phase showing that this is where Bremsstrahlung emission is most efficient.

In the next section we quantify the contribution of each of these wind phases to the total X-ray luminosity of the galaxy.

\subsection{Wind mixing phases} \label{sec:tracer}

\begin{figure}
    \centering
    \includegraphics[width=0.48\textwidth]{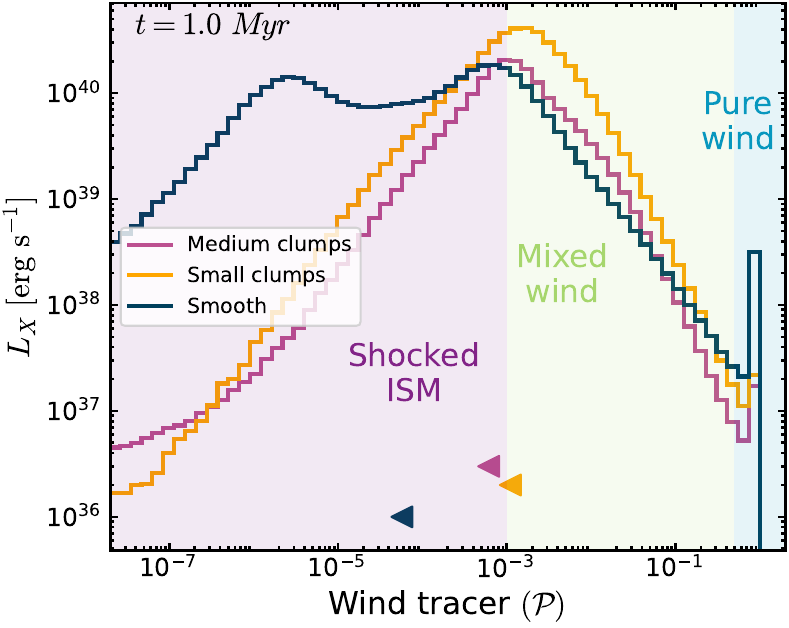}
    \caption{Histogram of X-ray luminosity as a function of the wind tracer ($\mathcal{P}$). Each bin shows the sum of the X-ray emission within it. We split the wind into pure, mixed and shocked ISM. The pink and yellow lines show the results from an initially clumpy setup, with {\LmaxEq{=}{170}} and {\LmaxEq{=}{40}} respectively. The blue line shows a smooth setup. The triangle points show the wind tracer value at which 50\% of the emission is produced. We can see that the smooth case is dominated by low-$\mathcal{P}$ values (shocked ISM), whereas the clumpy setup has a larger relative contribution from the mixed phase.}
    \label{fig:tracer}
\end{figure}

In Figure \ref{fig:tracer} we show a histogram of the estimated X-ray emission as a function of the passively advected wind tracer. Each bin shows the sum of the Bremsstrahlung emission within that bin. The pink line shows the fiducial simulation with medium-sized initial clumps (fiducial; \LmaxEq{=}{170}), the orange line shows the results from the setup with small initial clumps (\LmaxEq{=}{40}), and the blue line shows an initially homogeneous disc (we omit the large clump case for clarity, as the results were similar to the fiducial case). We split the wind tracer into three regimes, as described in the previous section. The triangular points show the wind tracer value at which 50\% of the X-ray luminosity is produced.


Considering first a wind propagating in a smooth medium (blue line), \cite{Nims2015} used an analytic model, based on the work of \cite{Faucher-Giguere2012}, to estimate the Bremsstrahlung emission, assuming an isotropic wind, and a spherically-symmetric, homogeneous ambient medium. They argued that the shocked ambient medium would contribute more to the Bremsstrahlung emission than the shocked wind phase analysed in \cite{Faucher-Giguere2012}. The blue line in Figure \ref{fig:tracer} shows our results when using our smooth disc. Our results concur with \cite{Nims2015} that the emission from the shocked wind (`pure wind'; $\mathcal{P}>0.5$) is subdominant compared to the shocked ambient medium (`shocked ISM'; $10^{-3}<\mathcal{P}<0.5$). Although the shocked wind is hot ({\TEq{\approx}{9}}), its low density results in weak X-ray emission. We find that most of the X-ray emission from the smooth disc is from gas with $\mathcal{P}\lesssim 10^{-3}$, representing gas that has barely mixed with the AGN wind. The dark blue triangle shows that 50\% of the X-ray luminosity is produced by gas with $\mathcal{P}<10^{-4}$ showing the emission is dominated by the unmixed, shocked phase.

However, our results differ from the findings of \cite{Nims2015} and our smooth model when we consider a clumpy medium, shown in pink (medium clumps) and orange (small clumps) in Figure \ref{fig:tracer}. Firstly, we can see that the emission from the pure wind is even lower than in the smooth case, possibly because the hot wind can flow out through the low density regions, causing the reverse shock to occur when the wind density is lower and resulting in less thermalisation \citep[see, e.g.,][]{Faucher-Giguere2012,Costa2020}.
The emission peaks for intermediate values of $\mathcal{P}$ -- this is ISM gas which has been ablated from the cold clumps and mixed with the injected wind (see bottom panels in Figure \ref{fig:overview}). The emission declines again for low wind tracer values, unlike in the smooth case which shows strong emission in the shocked ISM phase. This shows that when the ISM is arranged in a clumpy structure, it is much more difficult for shocks to cause it to radiate. This confirms what was postulated by \cite{Nims2015}, who predicted that an AGN wind would struggle to shock-heat dense clumps to high enough temperatures for luminous Bremsstrahlung emission. However, we find this is partly compensated for by enhanced mixing between the ISM and wind, resulting in more emission for the mixed wind phase in the clumpy case.

Additionally, we find that the size of the initial clumps plays a role in the resulting emission. We find that smaller initial clumps (\LmaxEq{=}{40}; orange line) have more emission across all values of $\mathcal{P}$ than the medium clumps (\LmaxEq{=}{170}; pink line). We didn't find any significant difference between the medium and large (\LmaxEq{=}{330}) clumps, so we omit the large clump runs for clarity (we show results for all the cloud sizes in Appendix \ref{sec:ap:params}). The enhancement in emission in the mixed wind phase for the small clumps could be due to a greater surface area of the ISM for the wind to interact with, resulting in more mixing between the hot wind and the cold ISM. Additionally, the slightly higher emission in the shocked ISM regime could be because the more tightly spaced clumps could be more efficient at trapping the expanding outflow, resulting in less wind venting and more shock-heating. This is supported by the result found in \citetalias{Ward2024} that small clumps resulted in a higher mass outflow rate in the hot gas phase.

We note that the exact boundary value between mixed wind and shocked ISM phase is an arbitrary choice. The value we have chosen ($\mathcal{P}_{\rm{mix}}\approx 10^{-3}$) represents a mass fraction of $<0.1\%$ of the cell being composed of mass initially injected by the wind for `shocked' or unmixed ISM. From Figure \ref{fig:tracer}, we can see that the clumpy case has far less X-ray emission at low $\mathcal{P}$ than the smooth case, showing the emission from clumpy discs is dominated by mixed gas. This broad conclusion is insensitive to the exact choice for $\mathcal{P}_{\rm{mix}}$.

\begin{figure*}
    \centering
    \includegraphics[width=0.98\textwidth]{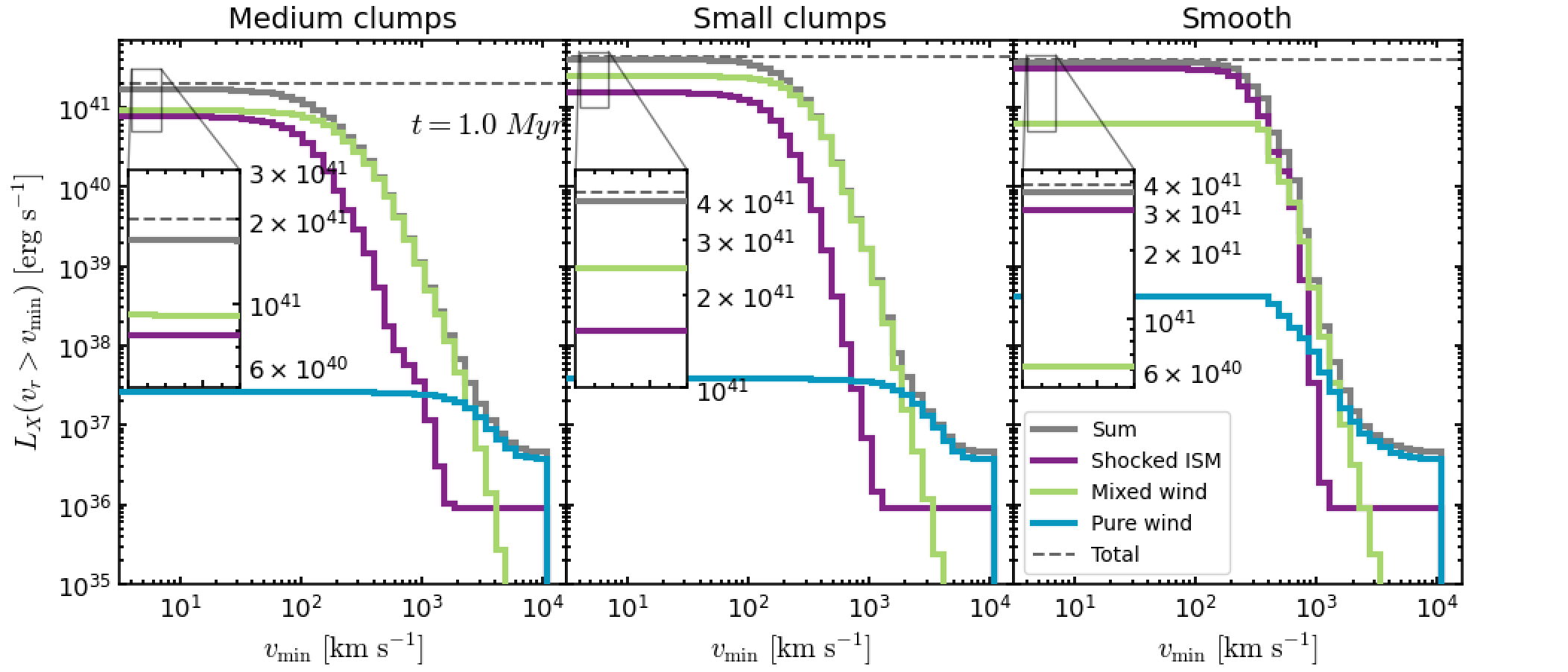}
    \caption{Cumulative total Bremsstrahlung emission as a function of outflow velocity, split into the wind phases as presented in Figure \ref{fig:tracer} with the mixed-ISM boundary set at $\mathcal{P}_{\rm{mix}}=10^{-3}$. The horizontal dashed line shows the emission over the whole velocity space (i.e., including the static background). We show (from left to right) the results from our medium clumps (fiducial), small clumps and smooth simulations. In the two clumpy cases, the pure wind dominates at the fastest velocities {\vrEq{\gtrsim}{3000}} (i.e., in the centre near injection and high-velocity vents) but this is negligible overall. Gas that has significantly mixed with the wind dominates from {\vrEq{\approx}{100-3000}}. Overall, the shocked ISM and the mixed wind have roughly equal contributions to the total X-ray luminosity for the clumpy case. However, in the smooth case the emission from shocked, unmixed ISM gas ($\mathcal{P}_{\rm{mix}}\lesssim 10^{-3}$) dominates the other sources.}
    \label{fig:cumul_LX}
\end{figure*}

\subsection{Total X-ray contribution by wind phase} \label{sec:cumul}

In Figure \ref{fig:cumul_LX} we show the total Bremsstrahlung luminosity, summed across all cells in the box at {\tEq{=}{1}}. We show the cumulative emission as a function of radial velocity, split by wind phase (see Section \ref{sec:tracer}). The left panel shows our fiducial simulation with medium clump sizes (\LmaxEq{=}{170}), the middle panel shows the simulation with small clump sizes (\LmaxEq{=}{40}), and the right panel shows an initially inhomogeneous setup.

For the medium clumps (left panel), we can see that the highest velocity gas is all contained in the pure wind phase (blue), starting at the injection velocity of \vrEq{=}{10^4}. However, this makes a small overall contribution to the overall X-ray luminosity ($L_X\approx 3\times10^{37}\ \rm{erg\ s^{-1}}$) and there is no gas moving slower than \vrEq{\lesssim}{1000} in this phase emitting any X-rays. At a radial velocity of \vrEq{\lesssim}{3000}, the mixed wind phase starts dominating over the pure wind before flattening off at \vrEq{\approx}{100}. At lower velocities, the shocked ISM phase has similar X-ray emission to the mixed wind. The shocked ISM phase is made up of gas that has been driven in front of the wind shock without mixing with it and cold clumps that have been heated but not destroyed by the wind. This phase is mostly moving at \vrEq{\approx}{100-500}. The inset shows that the total X-ray emission from the shocked ISM and mixed wind phases is similar, although the mixed phase has a slightly higher total of $L_X\approx 9\times10^{40}\ \rm{erg\ s^{-1}}$ compared to $L_X\approx 8\times10^{40}\ \rm{erg\ s^{-1}}$ in the shocked phase. This gives a total emission in the fiducial simulation of $L_{\rm{X}}=1.8 \times 10^{41}\ \rm{erg\ s^{-1}}$.

In the case with small initial clumps (middle panel of Figure \ref{fig:cumul_LX}), we can see a similar distribution to the fiducial run. However, the total luminosity is higher ($L_{\rm{X}}=4.0 \times 10^{41}\ \rm{erg\ s^{-1}}$) which is driven by increased emission in both the shocked and mixed phases. The increase in the mixed phase shows that the wind is able to interact with a greater amount of cold gas, possibly due to the higher surface area of the smaller clumps. In \citetalias{Ward2024}, we showed that the velocity distribution of the tightly clumped case started to tend towards that of the smooth case, suggesting the outflow was being more efficiently trapped by the smaller clumps. This could also be the reason for the elevated contribution from the shocked phase.

In the rightmost panel we plot the results for the smooth disc. We can see that the pure wind contribution is higher than in the clumpy case, however, it is still subdominant to the mixed and shocked phases, as also predicted by earlier works \citep[e.g.,][]{Faucher-Giguere2012,Nims2015}. The emission from the shocked phase rises more steeply than in the clumpy case -- i.e., there is a narrower velocity range where most of the emission is occurring. The total emission starts to flatten out at \vrEq{\approx}{300} compared to \vrEq{\approx}{100} in the clumpy case. This is consistent with our findings in \citetalias{Ward2024}, where in the smooth case the bulk of the mass of the outflowing material had a characteristic velocity of \vrEq{\approx}{400}, whereas in the clumpy case, there was a wider range of velocities (\vrEq{\approx}{10-300}). The total luminosity in the smooth run is $L_{\rm{X}}=3.8 \times 10^{41}\ \rm{erg\ s^{-1}}$; around double the value of the medium clumps, and just less than the small clumps. This demonstrates that the enhanced X-ray emission from wind-ISM mixing somewhat compensate the lack of emission from the shocked ISM in the clumpy case (see Figure \ref{fig:tracer}).

The grey dashed line in all three panels of Figure \ref{fig:cumul_LX} shows the X-ray emission summed over the whole simulation and the grey solid histogram shows the sum of the three outflow phases cumulative emission by radial velocity. In the fiducial case, the sum of the three outflow phases (solid grey line) is lower by around $2\times 10^{40}\ \rm{erg\ s^{-1}}$ (dashed grey line), corresponding to $\lesssim 10\%$ of the total X-ray emission. This difference comes from gas with \vrEq{\lesssim}{5} associated to the static, hot background. To confirm this, we calculated the X-ray emission from our no-AGN control run which summed to $L_{\rm{X,bkg}} =2.5\times 10^{40}\ \rm{erg\ s^{-1}}$. This emission would be classified as `shocked ISM' according to our wind tracer definitions, which further strengthens our conclusion that the emission from the outflow itself is dominated by the mixed phase. For the fiducial run of {\LAGNEq{=}{45}}, this background emission is fairly negligible, but it becomes more dominant for lower luminosity simulations. For this reason, in Sections \ref{sec:time_evo} \& \ref{sec:scaling}, we subtract this background value from the total emission to uncover the underlying trends.

We note that cloud ablation and turbulent mixing are problems with a high degree of sensitivity to the numerical resolution. In Appendix~\ref{sec:ap:res}, we calculate the Bremsstrahlung emission in each of these wind phases for runs with mass resolutions \MtargetEq{=}{10,100\ \&\ 1000} and discuss the impact of numerical convergence on our results. We find that the total Bremsstrahlung emission is reasonably well-converged at our fiducial resolution of \MtargetEq{=}{100}, although we note that the contribution from the mixed phase increases, and the shocked phase decreases, as the resolution increases. This further strengthens our finding that it is the wind-ISM mixing that is driving the bulk of the X-ray emission in our simulation.

\subsection{Radial evolution}

\begin{figure}
    \centering
    \includegraphics[width=0.48\textwidth]{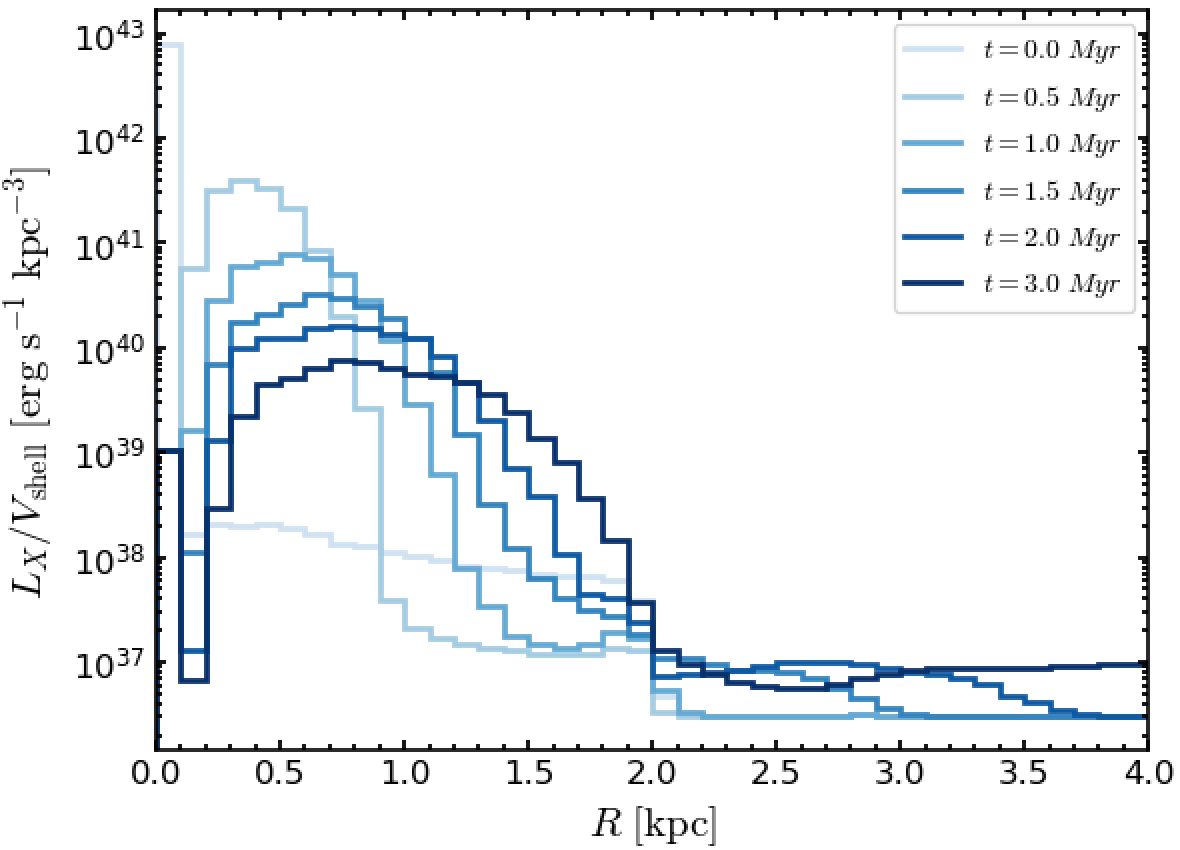}
    \caption{The radial evolution of the X-ray producing outflow. We show the X-ray luminosity in increasing radial shells, normalised by the volume of the shell. The peak of the outflow decreases and broadens with time, but still shows strong emission at radii $R<2\ \rm{kpc}$. There is some larger-scale emission in the halo ($R>2\ \rm{kpc}$), but this is much fainter than the outflow within the disc, although we not this may be due to our simplistic CGM model.}
    \label{fig:timerad}
\end{figure}

Figure \ref{fig:timerad} shows the radial evolution of the X-ray producing gas for a range of times up to \tEq{=}{3} in the fiducial simulations. The galaxy is binned in spherical shells of radius $R$ and the luminosity is divided by the volume of each shell. Darker lines show later times. The X-ray emission starts strongly peaked, before widening and steeply reducing in peak flux. The peak of the emission expands with a velocity of $v \approx 250 \ \rm{km\ s^{-1}}$, demonstrating it is dominated by the equatorial outflow rather than the faster-moving polar outflow. By \tEq{=}{3}, the emission has broadened significantly across a width of $R\approx 1.5\ \rm{kpc}$. As shown in Figure \ref{fig:overview}, in our setup, most emission is formed in the disc, rather than in the halo bubble. This is because the overall emission is dominated by ISM gas that has strongly mixed with the wind, rather than the wind material itself (Figures \ref{fig:tracer} \& \ref{fig:cumul_LX}). Once the outflow breaks out of the disc in the polar direction, there is less material to mix with, due to the halo's low density. We can see in Figure \ref{fig:timerad} that at large radii ($R\gtrsim2\ \rm{kpc}$) the emission is significantly weaker than within the disc, although this may be driven by our simplistic halo density model, as discussed in Sections \ref{sec:sum_fig} \& \ref{sec:cgm}.

Our prediction for X-ray emission on extended ($D \approx 3\ \rm{kpc}$) scales within the disc suggests that is could be distinguished from the point-source emission from the AGN itself. We discuss the detectability of this emission further in Section \ref{sec:observations}.

\subsection{Time Evolution} \label{sec:time_evo}

\begin{figure}
    \centering
    \includegraphics[width=0.48\textwidth]{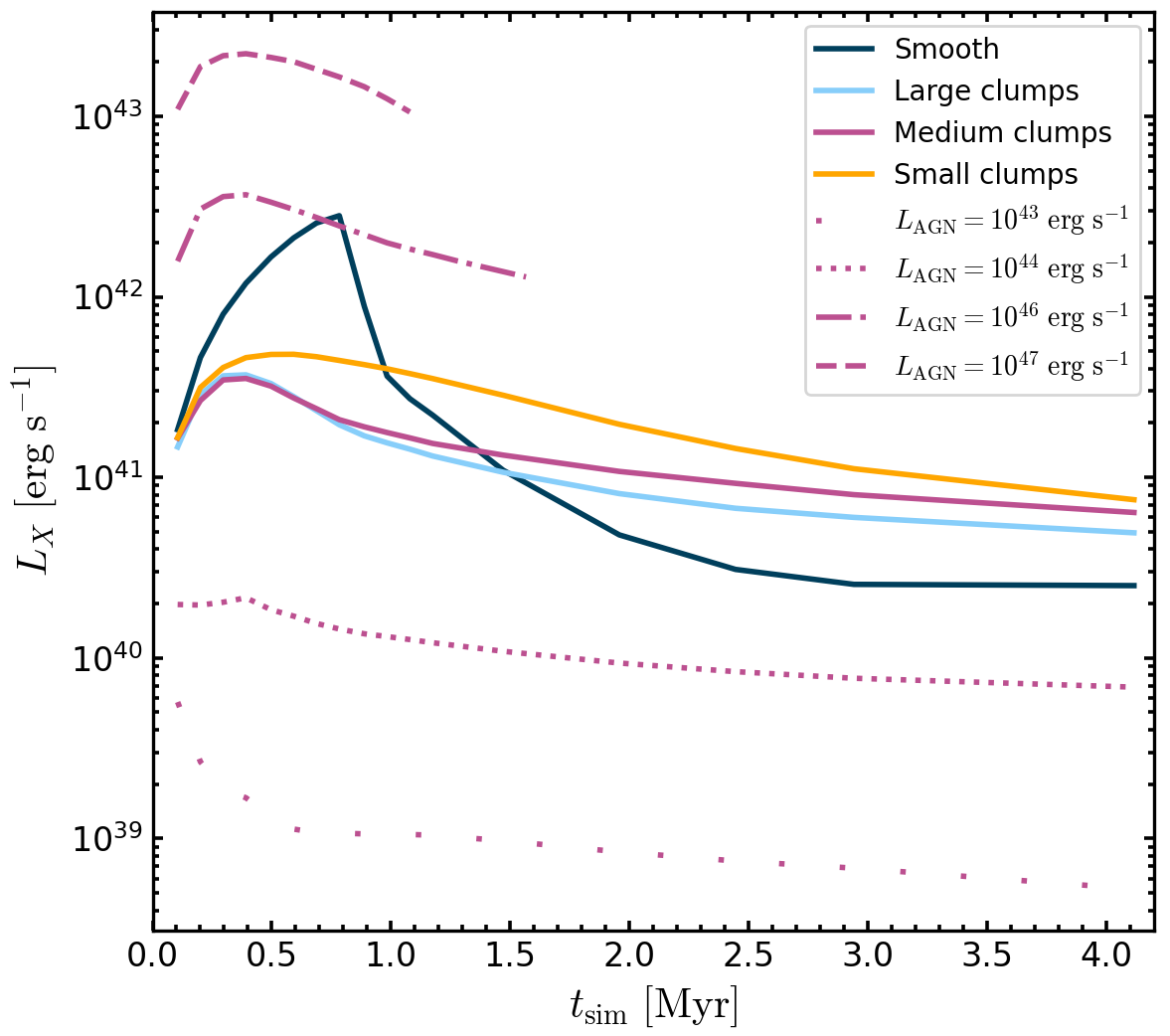}
    \caption{The time evolution of the X-ray luminosity from Bremsstrahlung emission. The contribution from the static background ($2.5\times 10^{40}\ \rm{erg\ s^{-1}}$) has been subtracted to reveal the underlying trends. The pink lines show the disc with medium-sized clumps at a range of AGN luminosities (as shown by variable line-style). The dark blue lines show an initially smooth disc at \LAGNEq{=}{45}, and the light blue/orange lines show initially large/small clump sizes.}
    \label{fig:time_param}
\end{figure}

Figure \ref{fig:time_param} shows the time evolution for the Bremsstrahlung X-ray luminosity from our simulations. We show results for a range of AGN luminosities (\LAGNEq{=}{43-47}; dashed/dotted lines) in discs with our fiducial clumpiness (\LmaxEq{=}{170}; `medium' clumps). For an AGN luminosity of \LAGNEq{=}{45}, we also show results for variations in clumpiness, including large clumps (\LmaxEq{=}{330}; light blue), small clumps (\LmaxEq{=}{40}; orange), and the smooth setup (dark blue). We subtract the background emission of $L_{\rm{X,bkg}} = 2.5\times 10^{40}\ \rm{erg\ s^{-1}}$ from these values (see Section \ref{sec:cumul}).

In our fiducial simulation (solid pink line), the luminosity rises to a peak of $L_{\rm{X}}\approx 2.5 \times 10^{41}\ \rm{erg\ s^{-1}}$ at \tEq{\approx}{0.3} before declining slowly to $L_{\rm{X}}  \approx 6 \times 10^{40}\ \rm{erg\ s^{-1}}$ by \tEq{\approx}{4}. This peak represents the breakout time for the hot wind, after which the wind can more easily vent out of the low density channels resulting in less mixing with the denser ISM. The large clump case (light blue) is very similar to the fiducial at \tEq{\lesssim}{1} but then declines slightly faster to $L_{\rm{X}}  \approx 5 \times 10^{40}\ \rm{erg\ s^{-1}}$. The small clumps (orange) show a higher and slightly later peak in emission of $L_{\rm{X}}\approx 5 \times 10^{41}\ \rm{erg\ s^{-1}}$ and declines less rapidly than the medium and large clumps. This could be due to the smaller clumps better trapping the hot wind \citep{Ward2024}, resulting in a longer mixing phase before the venting channels are clearer. 

Our results for the clumpy ISM bear some similarity with the time evolution of the X-rays produced by the jet-ISM interaction in \cite{Sutherland2007}. Although their model uses a jet, their sequence of `flood-and-channel', disc breakout, and longer-lasting radiative shocks in the ISM, seems to qualitatively match the steep rise, broad peak, and long tail of our time evolution.

Conversely, the smooth setup (dark blue lines) shows a rapid increase in luminosity to a peak of $L_{\rm{X}}\approx 3 \times 10^{42}\ \rm{erg\ s^{-1}}$ at \tEq{\approx}{0.8} before swiftly declining. This corresponds to the `post-shock-cooling' time discussed in \citetalias{Ward2024}. At this time, the amount of cold gas in the outflow rapidly increases due to efficient cooling, but Figure \ref{fig:time_param} shows that this results in an equally sudden drop in X-ray emission due to much less hot gas in the outflow. By \tEq{\approx}{1.5}, the smooth case has the lowest emission of any of the \LAGNEq{=}{45} simulations.

In Figure \ref{fig:time_param} we also show the time evolution for a range of AGN luminosities.\footnote{The highest AGN luminosity runs (\LAGNEq{=}{46-47}) are evolved to shorter times due to computational and box-size constraints.} The highest luminosity AGN show the brightest Bremsstrahlung emission, with the X-ray luminosities for \LAGNEq{\geq}{44} all showing a similar shape; rising to a peak at \tEq{\approx}{0.3} before declining slowly, although the peak in the \LAGNEq{=}{44} track is quite small. However, the lowest AGN luminosity (\LAGNEq{=}{43}) shows a different trend. The emission peaks at the start before rapidly declining to $L_{\rm{X}}\approx 10^{39}\ \rm{erg\ s^{-1}}$ by \tEq{\approx}{0.5}. This could suggest the AGN is not powerful enough for the outflow to impact and mix with a significant amount of ISM gas. For example, at \tEq{=}{0.5}, the \LAGNEq{=}{43} simulation has $M_{\rm{mix}}=10^{5.3}\ \rm{M_\odot}$ in the mixed gas phase, compared to $M_{\rm{mix}}=10^{7.1}\ \rm{M_\odot}$ when \LAGNEq{=}{45} -- a two dex difference roughly matching the variation in Bremsstrahlung emission between these two simulations. 

We can see that the simulations with different {\LAGN} have roughly equally-spaced increases in X-ray luminosity. In Section \ref{sec:scaling}, we will investigate the correlation between the AGN luminosity and resulting X-ray emission.

\begin{figure}
    \centering
    \includegraphics[width=0.49\textwidth]{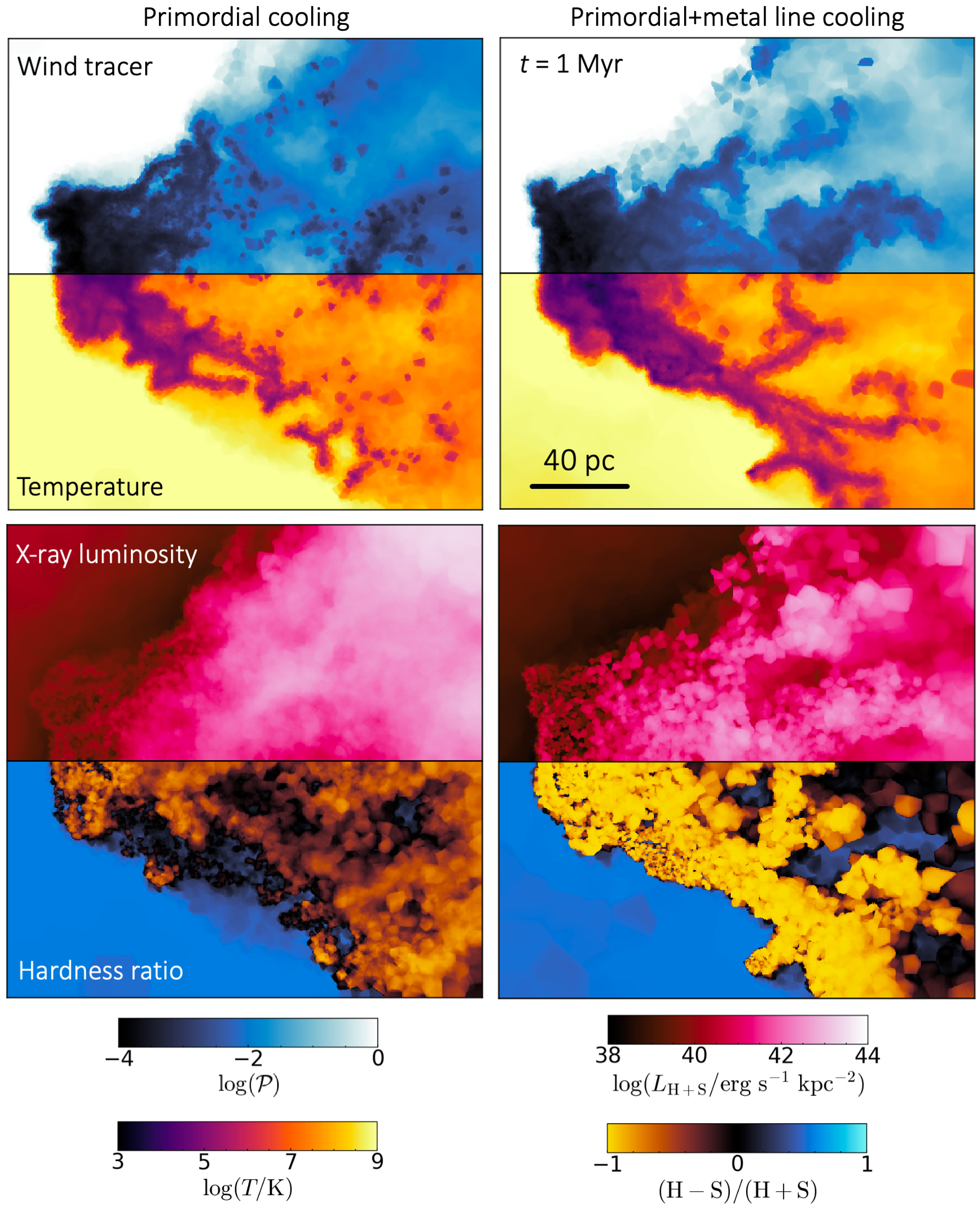}
    \caption{A zoomed-in view of a cold cloud from the simulations with primordial cooling (left column) and including metal line cooling (right column), showing wind tracer density, temperature, X-ray luminosity, and hardness ratio. Luminosity and hardness ratio are calculated using the X-ray bands $L_{\rm{S}}=(0.5-2)\ \rm{keV}$ and $L_{\rm{H}}=(2-10)\ \rm{keV}$. The AGN wind is approaching from the left and the cloud is projected through a depth of 50 pc. The resulting mixing structure of the cloud in both simulations is similar, although the increased cooling in the metal line run results in a more filamentary structure in the cold phase and brighter and softer X-ray emission due to the high-ionisation iron line complex at $E\simeq1\ \rm{keV}$.}
    \label{fig:coolingcloudZ}
\end{figure}

\subsection{Metal line cooling} \label{sec:results line cooling}

As described in Section \ref{sec:line cooling}, this work builds on \citetalias{Ward2024} by including metal lines as an additional cooling channel. This was motivated in part by the need to include metal line emission in the X-ray calculations (see Section \ref{sec:spectra}), but also allowed us to investigate the effect of this additional cooling channel on the outflow.

To visualise the impact of metal line cooling on the structure of the outflow, Figure \ref{fig:coolingcloudZ} shows a zoom-in of a $d\simeq 40\ \rm{pc}$ cloud being impacted by the AGN wind from the left at a time of \tEq{=}{1}. The left-hand column shows the simulation with primordial cooling only and the right-hand column shows the impact of also including metal-line cooling. From top to bottom, the panels show the AGN wind tracer, the temperature, the X-ray luminosity from $0.5-10\ \rm{keV}$, and the hardness ratio defined as $\rm{HR}=(L_{2-10\rm{keV}}-L_{0.5-2\rm{keV}}) / (L_{0.5-2\rm{keV}}+L_{2-10\rm{keV}})$. The X-ray luminosity is integrated over the box depth of $50\ \rm{pc}$, the temperature and wind tracer are density-weighted averages, and the hardness ratio is weighted by the X-ray emission. The X-ray emission in the metal line cooling case includes line emission, as described in the next section. In general, the shape of the cloud and temperature structure around the mixing layers are similar between the two cases. However, in the simulation with metal line cooling, the cold cloud is slightly larger, and the stripped cold clouds in the tail have a more filamentary structure compared to the primordial-only case, showing the increased cooling rate. The simulation including metal line cooling also shows enhanced X-ray emission in the tail, and the bottom-right panel shows that this emission is very soft. This demonstrates the impact of the high-excitation iron line complex at $E\simeq1\ \rm{keV}$ seen in Figure \ref{fig:spectra} which contributes significantly to the soft band.

In Appendix \ref{sec:ap:Zcooling} we discuss further how the inclusion of metal line cooling impacts the outflow structure. In Figure \ref{fig:ap:vr_cooling} we show that including metal line cooling results in slightly faster moving cool gas (\TEq{\leq}{4.5}), up to \vrEq{\simeq}{700}. In Figure \ref{fig:ap:energy_driving}, we replicate our result from Figure 10 in \cite{Ward2024} where we investigated the energy-driving of the outflow. We find that, despite the increased cooling, there is only a slight decrease in the momentum flux from \pEq{\simeq}{6} in the primordial-only case to \pEq{\simeq}{5} when including metal line cooling. This remains comfortably in the energy-driven regime (\pEq{>}{1}), demonstrating that the outflow is behaving qualitatively the same with either cooling model.

\subsection{Emission from metal lines} \label{sec:spectra}

\begin{figure}
    \centering
    \includegraphics[width=0.45\textwidth]{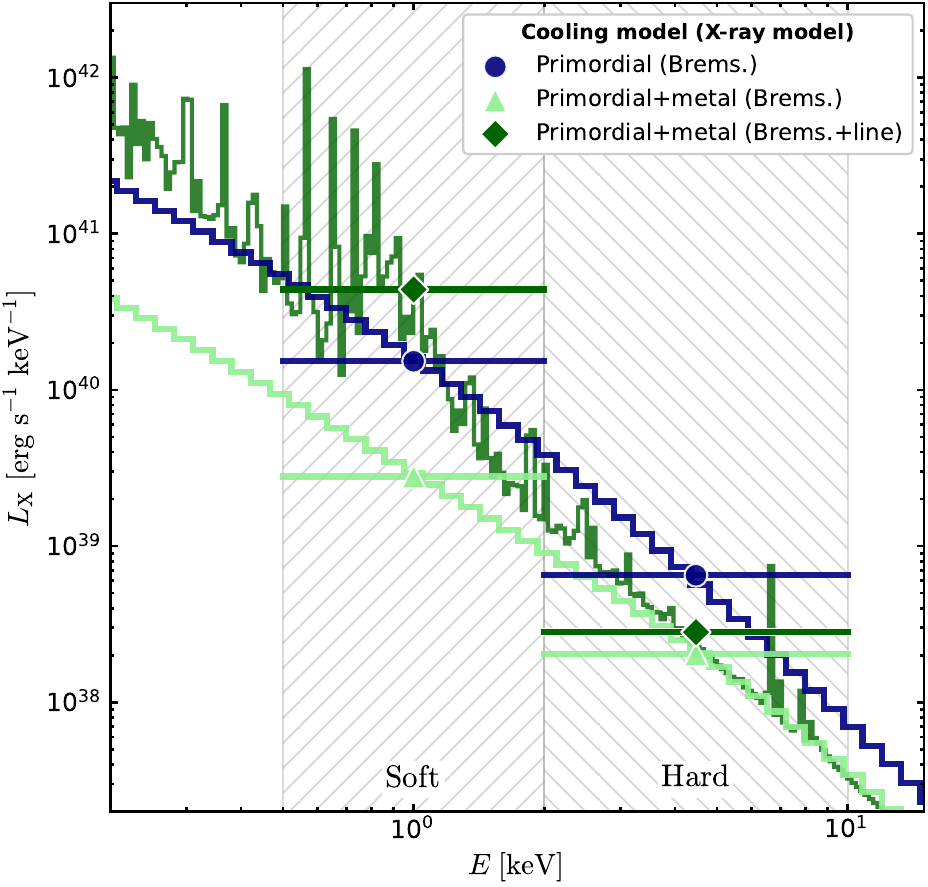}
    \caption{Rest-frame X-ray spectra from the fiducial simulation at \tEq{=}{1}, after background subtraction. The dark blue line shows the spectrum of the Bremsstrahlung emission from the run with primordial cooling and the light blue lines shows the Bremsstrahlung emission when metal line cooling is included. The green line shows the spectra from the line cooling simulation when metal line emission is included. This significantly boosts the soft emission due to a complex of high-excitation iron lines at around $E=1\ \rm{keV}$.}
    \label{fig:spectra}
\end{figure}

\begin{figure}
    \centering
    \includegraphics[width=0.33\textwidth]{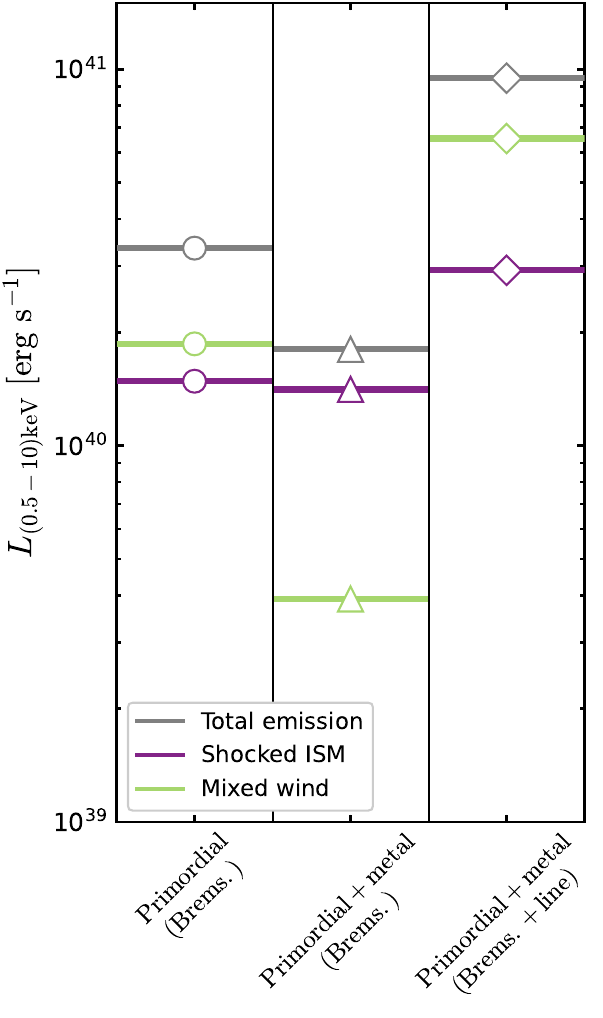}
    \caption{The contribution of wind mixing phases to the total X-ray luminosity for three different cooling and X-ray emission models: the left column (circles) shows the simulation with only primordial cooling and with the X-ray emission estimated using the Bremsstrahlung approximation; the middle column (triangles) shows the simulation with both primordial and metal line cooling, but only considering the Bremsstrahlung contribution to the X-rays; and the right column (diamonds) shows the primordial+metal line simulation with both Bremsstrahlung and metal line emission contributions to the X-ray luminosity. The colour of the lines show the mixing phase of the gas, with green showing the mixed phase, purple showing the poorly-mixed shocked ISM, and grey showing the total emission in all phases (the pure wind emission is negligible so has been omitted for clarity).}
    \label{fig:compare_LZ}
\end{figure}

In Sections \ref{sec:sum_fig}--\ref{sec:time_evo}, we used the Bremsstrahlung approximation to estimate the X-ray emission. This was a valid assumption for our simulations that only included primordial cooling, and allowed us to compare against previous analytic work such as \cite{Nims2015}. However, to make a better comparison between our simulations and X-ray observations, we needed to include additional effects such as metal line cooling and radiative recombination which could increase the emission from the cooling gas in observable energy bands.

In Figure \ref{fig:spectra} we show the predicted rest-frame spectrum of the X-ray emission from our simulations, covering the soft ($0.5-2\ \rm{keV}$) and hard ($2-10\ \rm{keV}$) X-ray bands. The spectrum is calculated by numerically evaluating Equation \ref{eq:freq int} over small ($\Delta \log (E/\rm{\ keV})=0.01$) frequency increments. This just includes the ISM gas in our simulation, excluding any emission from the point-source AGN itself. As in earlier sections, we subtracted the emission from the hot background gas by computing the spectra in the same way for the non-AGN simulations run with and without metal-line cooling. 

The dark blue lines show the spectrum from just the Bremsstrahlung mechanism for the simulations with primordial cooling and the light green lines shows the result for the Bremsstrahlung emission in the simulations with primordial and metal-line cooling. We can see that the inclusion of line cooling reduces the Bremsstrahlung emission predicted by our simulations by a factor of 5-10 across the whole energy range. This is because gas with temperatures \TEq{\simeq}{5-7}, is now able to cool more efficiently, resulting in a lower total mass of gas staying at these temperatures and radiating X-rays.

We quantify the effect of metal line cooling on the mixing structure of the wind-cloud interaction in Figure \ref{fig:compare_LZ} where we show the integrated X-ray luminosity across the soft and hard bands ($0.5-10\ \rm{keV}$) for primordial cooling only (circles) and primordial plus metal line cooling (triangles/diamonds). The grey line shows the total luminosity, and this is further broken down into the contribution from the shocked ISM (purple) and mixed wind-ISM (green), according to the definitions in Section \ref{sec:tracer}. The pure wind makes a negligible contribution, so is omitted in this plot. 
We can see that switching on metal line cooling makes only a small difference to the shocked ISM emission, but significantly reduces the contribution from the mixed phase.
As we explored in Section \ref{sec:results line cooling}, the enhanced cooling in the temperature range \TEq{\simeq}{6-7} from the metal lines results in quicker cooling of this intermediate temperature gas onto the more pronounced cold filaments. This reduces the total mass at this temperature (also seen in Figure \ref{fig:ap:vr_cooling}) which decreases the integrated Bremsstrahlung emission from this wind phase, especially in the soft band. This supports the argument that it is this mixed phase at \TEq{\simeq}{6-7} which is affected most by line cooling.

However, metal line transitions also contribute significantly to the X-ray emission. In Figure \ref{fig:spectra} we show the simulated spectra including these line emission contributions in dark green. The inclusion of line emission has only a modest effect on the hard X-ray band ($2-10\ \rm{keV}$), although we can see a strong \ion{Fe}{XXV} line at 6.7 keV which is typical of collisionally-excited gas \citep{Fabbiano2022chapter}. The effect on the soft band is much greater, resulting in a $>1\ \rm{dex}$ increase in emission due to a complex of lines at $\simeq 1\ \rm{keV}$, mostly due to high-excitation iron (\ion{Fe}{XVII-XXIV}; \citealt{Gu2019}). The inclusion of emission from radiative recombination when using \textsc{PyAtomDB} to calculate the spectrum also boosts the continuum in the soft band.

These effects can be seen clearly in Figure \ref{fig:compare_LZ}, where the total X-ray emission is significantly boosted by the metal lines and recombination. This increase is driven by the mixed phase of the gas (green diamonds) which becomes the dominant source of the X-ray emission. This shows that, despite there being less overall gas in this temperature and density regime when line cooling is enabled, the additional emission from these metal lines more than compensates when considering the total X-ray emission. This strengthens the argument of this paper, which is that wind-ISM mixing dominates the X-ray emission in AGN outflows and that this emission is a consequence of the cooling required for the cold outflow to survive \citep{Ward2024}.

\subsection{Scaling relations between AGN and X-ray luminosities} \label{sec:scaling}

\begin{figure}
    \centering
    \includegraphics[width=0.49\textwidth]{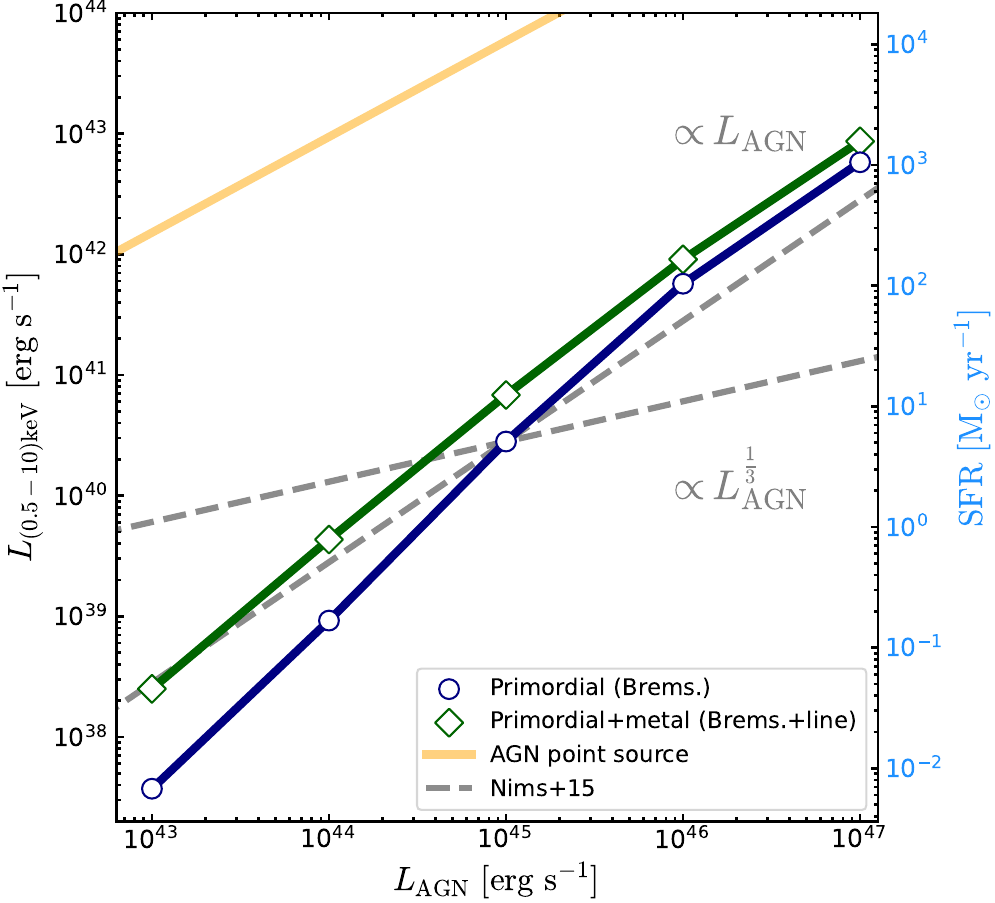}
    \caption{Correlations between AGN luminosity and the X-ray emission of various processes. In blue, we show the Bremsstrahlung emission from our simulations with primordial cooling at {\protect \tEq{=}{1}}. We compare this to two analytic predictions for this process from {\protect \cite{Nims2015}} . In green, we show the predicted scaling for our simulations with metal line cooling and emission. We compare the predicted emission from our simulations with two other sources of X-rays: the point-source AGN (shown in orange; {\protect \citealt{Marconi2004}}), and from star formation (shown on the right-side axis; {\protect \citealt{PereiraSantaella2011}}). Both of these components are integrated across the range $0.5-10\ \rm{keV}$. Although the point-source AGN dominates the X-ray emission in galaxies, our predicted mixing-driven X-rays may be detectable against the diffuse star formation component, if sufficiently resolved.}
    \label{fig:LXLbol}
\end{figure}

In Figure \ref{fig:LXLbol} we show the predicted scaling relation between the AGN luminosity and the resulting X-ray luminosity of the outflow.
In blue we show the Bremsstrahlung emission from our simulation with primordial cooling, and in dark green we show the simulations with primordial and metal line cooling, with both Bremsstrahlung and metal line X-ray emission. We show our results at a time of \tEq{=}{1}, which is just after the peak in emission shown in Figure \ref{fig:time_param}, and subtract the background emission (negligible for \LAGNEq{>}{44}, see Section \ref{sec:cumul}). We find a positive correlation between the AGN and X-ray luminosities across the whole parameter range of \LAGNEq{=}{43-47}.

We compare this result with theoretical predictions from \cite{Nims2015} who predicted a scaling between the AGN luminosity and Bremsstrahlung emission of $L_{\rm{X}} \propto L_{\rm{AGN}}^{1/3}$ in the case of inefficient cooling of the shocked ISM, or $L_{\rm{X}} \propto L_{\rm{AGN}}$ if the ISM can cool rapidly. We plot these two cases in grey dotted lines and set the arbitrary normalisation of these tracks to the luminosity from the fiducial simulation (\LAGNEq{=}{45}). Our results match the rapid cooling case, with a gradient of just over unity across the whole luminosity range, with perhaps a slight tail off for the brightest AGN. This scaling of $L_{\rm{X}} \propto L_{\rm{AGN}}$ can also be justified as follows: in the AGN wind model used in this work \citep{Costa2020}, the density of the wind is proportional to the AGN luminosity: $n_{\rm{w}} \propto L_{\rm{AGN}}$. As the wind mixes with the ISM, the density of this mixed material becomes $n_{\rm{mix}} \approx \sqrt{n_{\rm{w}}\ n_{\rm{ISM}}}$. From the Bremsstrahlung approximation given in Equation \ref{eq:brem}, the X-ray emission is thus given by:

\begin{equation}
    L_{\rm{X}} \propto n_{\rm{mix}}^2 \propto n_{\rm{w}}\ n_{\rm{ISM}} \propto L_{\rm{AGN}}\ n_{\rm{ISM}}
\end{equation}

which yields the linear relationship found in Figure \ref{fig:LXLbol}. This confirms that our outflow is in the rapid cooling regime predicted by \cite{Nims2015} for either of our cooling models.

\section{Discussion} \label{sec:discussion}

Now we have characterised the X-ray emission caused by wind-ISM interactions, the next question is: is this observable? In this section we will explore the relative luminosities of various sources of X-rays from galaxies and the spatial scales these act on, and discuss whether these mixing-driven X-rays could be detected with telescopes such as \textit{Chandra}.

\subsection{Other sources of X-rays}

The two major sources of X-rays in galaxies are star formation and the AGN itself. In Figure \ref{fig:LXLbol} we plot tracks showing the rough contribution of each of these to the X-ray luminosity to contextualise the magnitude of the emission predicted from the wind-ISM interactions.

Star formation produces X-rays mostly through X-ray binaries (XRBs). We use the relations described by \cite{PereiraSantaella2011} (their Eqs. 4 \& 5) for a sample of local galaxies to estimate the total X-ray luminosity expected from star formation,

\begin{equation}
    \frac{L_{\left(0.5-10\right)\rm{keV}}}{\rm{erg\ s^{-1}}} = 5.5\times 10^{39}\ \frac{\rm{SFR}}{\rm{M_\odot\ yr^{-1}}}
\end{equation}

This relation is shown as an additional scale on the right of Figure \ref{fig:LXLbol} in blue. At our fiducial AGN luminosity of \LAGNEq{=}{45}, the X-ray luminosity associated with the outflow of $L_{\rm{X}}=2\times 10^{41}\ \rm{erg\ s^{-1}}$ is equivalent to a star formation rate of around $\rm{SFR} \approx 15\ \rm{M_\odot\ yr^{-1}}$ which is a modest star formation rate for local galaxies.  For an outflow driven by an AGN with luminosity of \LAGNEq{=}{46}, the comparable star formation is $\rm{SFR} \approx 200 \rm{M_\odot\ yr^{-1}}$ which exceeds all but the most luminous starbursts in local quasar-hosting galaxies \citep[e.g.,][]{Stanley2017,Zhuang2021}. Additionally, the mixing-driven X-ray emission is likely to be more centrally concentrated than the star formation, making it a dominant contributor to the total X-ray luminosity for radii of a few kiloparsecs. 

This comparison suggests that, for quasars in the local Universe with typical star formation rates, we expect the emission from wind-ISM interactions to be equal to or even more luminous than that produced by star formation. This suggests that observed extended X-ray emission around quasars could be dominated by wind-driven mixing, rather than XRBs. Further work on spectral modelling of these two mechanisms could help distinguish between them in observed systems.

The orange line in Figure \ref{fig:LXLbol} shows the expected X-ray emission from the AGN itself, based on summing the soft and hard X-ray-bolometric corrections of \cite{Marconi2004}. This emission is caused by photons that have been up-scattered from the accretion disc, resulting in highly luminous, centrally concentrated X-ray emission. This X-ray source dominates over both the star formation and mixing-driven emission, being over two orders of magnitude higher than the line emission case shown by the green line. However, these X-rays are produced on the scale of the AGN accretion disc which is unresolvable, resulting in point-source emission. Therefore, if the spatial resolution of the observation is sufficient to resolve the mixing-driven emission, it should be able to be disentangled from the point-source accretion disc. We note that this is an optimistic assumption, as nuclear X-ray emission from the AGN may scatter into the host galaxy, contributing to the extended emission (see Section \ref{sec:photoionisation}). Conversely, in the case of Compton-thick Type 2 AGN, the nuclear emission is likely obscured by the dusty torus, creating a `natural coronagraph' \citep{Fabbiano2022chapter} which may make the detection of the mixing emission easier. As a simplifying assumption, we treat the AGN as a point source in this proof-of-concept analysis of the observability of outflow-driven X-ray emission.

We will now use these assumptions to discuss the maximum distance of a galaxy for this region to be spatially resolved.

\begin{figure}
    \centering
    \includegraphics[width=0.35\textwidth]{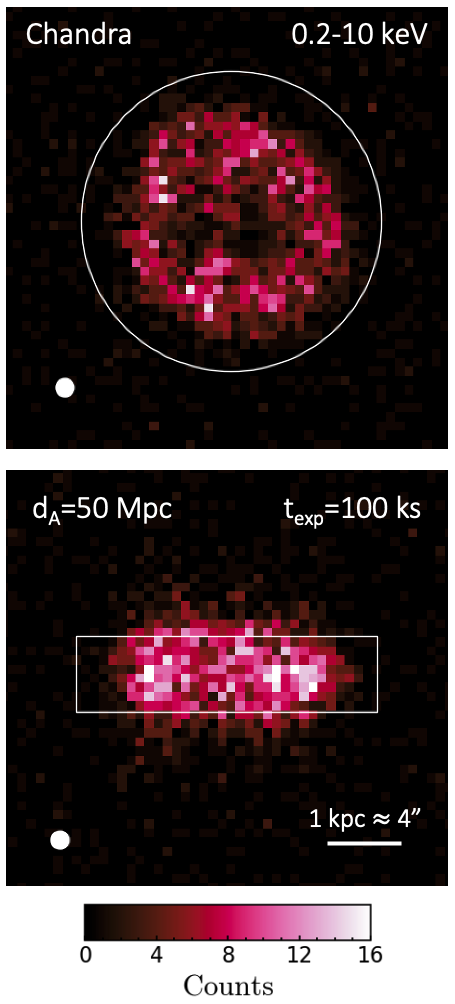}
    \caption{Mock observations of our predicted X-ray emission using \textit{Chandra-ACIS} for our galaxy at \tEq{=}{3} with an exposure time of $t_{\rm{exp}}=100\ \rm{ks}$. The top panel is face-on to the disc, and the bottom panel is side-on. The white outlines show the dimensions of the initial disc. At the angular diameter distance shown ($d_{\rm{A}}=50\ \rm{Mpc}$; $z\approx0.011$), 1 kpc is approximately $4"$ (scale bar in the lower right). We show the PSF of the \textit{Chandra-ACIS} in the lower left ($\rm{FWHM}=0.5"$).}
    \label{fig:chandra}
\end{figure}

\subsection{Observations with X-ray telescopes} \label{sec:observations}

In this section we discuss whether the prediction emission from wind-ISM mixing could be observable, using the \textit{Chandra} X-ray observatory as a baseline instrument. \textit{Chandra} boasts a high spatial resolution of $0.5"$, which has allowed detailed mapping of AGN outflows in several nearby galaxies (e.g., \citealt{Croston2008,Feruglio2013,Greene2014,DiGesu2017,Lansbury2018}, also see Table 1 in \citealt{Fabbiano2022chapter} for more examples).

In Figure \ref{fig:chandra} we show the photon counts per pixel expected from our simulation, forward modelled using the \textsc{PyXSIM}\footnote{\url{https://hea-www.cfa.harvard.edu/~jzuhone/pyxsim}} Python package (\citealt{ZuHone2016}, based on the \textsc{PHOX} code; \citealt{Biffi2012,Biffi2013}) which is then fed to the \textsc{SOXS}\footnote{\url{https://hea-www.cfa.harvard.edu/soxs}} instrument simulator \citep{ZuHone2023} to model \textit{Chandra} observations. We place our galaxy at a simulation time of \tEq{=}{3} at an angular diameter distance of $d_{\rm{A}} = 50\ \rm{Mpc}$ ($z\simeq 0.011$; representing a fairly local galaxy) and show the results for a wide X-ray band of $0.2-10\ \rm{keV}$ with an exposure time of $t_{\rm{exp}}=100\ \rm{ks}$. At this distance, 1 kpc spans 4" (shown by the scale bar in the lower right). We use the in-built SOXS model for the ACIS (Advanced CCD Imaging Spectrometer) instrument on \textit{Chandra}, and show the point-spread function (PSF) of this instrument in the lower left ($\rm{FWHM}=0.5"$). This is an idealised image, showing just the X-ray emission due to wind-ISM interactions presented in this work.


Firstly, we can see the X-ray emission is concentrated within the disc of the galaxy (shown in white outlines; $D=4\ \rm{kpc}$). As discussed in Section \ref{sec:cumul}, in a clumpy medium, the X-ray emission from the gas is dominated by mixed material being stripped from cold clouds. This makes the equatorial outflow X-ray bright, and shows us that this mixing-driven emission is a good tracer for the amount of ISM the wind has interacted with. The side-on view (lower panel of Figure \ref{fig:chandra}) also shows some chimneys extending out the top and bottom of the disc \citep[see e.g.,][]{Ponti2019,Ponti2021}. However, we don't see large X-ray bubbles extending into the halo of the galaxy, as has been seen in observational work \citep{Lansbury2018,Predehl2020}. We discuss this further in Section \ref{sec:caveats}.

As shown in Figure \ref{fig:LXLbol}, the dominant source of X-rays is expected to be the emission from the accretion disc. This emission is unresolved and acts a point source so could therefore be separated from the mixing-induced region if this component is well-resolved. At \tEq{=}{3}, the diameter of the mixed region is $D\approx 3\ \rm{kpc}$ (Figure \ref{fig:timerad}). We assume the structure to be sufficiently resolved at an angular size of three resolution elements, which for \textit{Chandra's} $\rm{FWHM}=0.5"$ PSF is $\theta=1.5"$. Therefore, for the $D= 3\ \rm{kpc}$ mixing region, we could resolve this emission at angular diameter distances of $d_{\rm{A}} \lesssim 410 \ \rm{Mpc}$, corresponding to a redshift of $z \lesssim 0.11$. However, we note that scattering could blur the nuclear emission beyond just a point-source, so this represents an optimistic scenario. Diffuse emission from star formation will also contribute to the total X-rays observed \citep[e.g.,][]{Lehmer2010}. However, as we have shown in Figure \ref{fig:LXLbol}, the X-ray emission from wind-ISM interactions should dominate over the XRB contribution in luminous quasars (\LAGNEq{\gtrsim}{45}), except for in the most extreme starbursts. 

Our model therefore predicts that, for local quasars ($z\lesssim0.1$), there should be an observable soft X-ray component extending for a few kpc beyond the central point source AGN. To investigate this prediction, we could use an observational sample such as the Quasar Feedback Survey (QFeedS; \citealt{Jarvis2021,Njeri2025}). QFeedS targets the multiphase outflows of a sample of $z\approx 0.05-0.2$ luminous quasars (\LAGNEq{\gtrsim}{45}) with typical star formation of $\rm{SFR} \approx 8-80\ \rm{M_\odot\ yr^{-1}}$. Our findings in this study suggest that these systems would be an ideal test-bed for studying the emission predicted from our model. For example, the Teacup galaxy (SDSS J1430+1339; \citealt{Harrison2015,RamosAlmeida2017,Audibert2023}), most famous for its large X-ray/radio bubble at 10 kpc scales, also shows a soft excess at $R\simeq 4\ \rm{kpc}$ scales which could be caused the wind-ISM interactions presented in this study.

\subsection{Model caveats and improvements} \label{sec:caveats}

The simulations presented in this work represent an experimental approach to understanding AGN-ISM interactions, and thus the model has several simplifications and caveats as described in \citetalias{Ward2024}. We will now briefly discuss two of the main limitations of this work that have the greatest implication for X-ray studies of AGN outflows.

\subsubsection{X-ray bubbles in the CGM} \label{sec:cgm}

Several studies have found large ($\gtrsim\ \rm{kpc}$) X-ray bubbles extending into the CGM of quasar host galaxies \citep[e.g.,][]{Croston2008,Greene2014,DiGesu2017,Lansbury2018}. Our own Milky Way also displays similar structures (the eROSITA bubbles; \citealt{Predehl2020}). However, our mock \textit{Chandra} observation in Figure \ref{fig:chandra} shows no such large-scale features, despite the relatively deep exposure time. We can see bubbles in Figure \ref{fig:overview}, but as mentioned in Section \ref{sec:sum_fig}, the low density and temperature contrast between these bubbles and the background results in weak X-ray surface brightness. This is in contrast to other simulations, such as \citep{Pillepich2021} who found eROSITA-like bubbles in IllustrisTNG50 using a wind-like model, and \cite{Jennings2025} who found bubbles using \textsc{Simba/Hyena}'s jet-like model. Our inability to detect these features is possibly due to our simplistic halo modelling -- we use a static, isobaric background with a constant density ($n_{\rm{bkg}}=10^{-2}\ \rm{cm^{-3}}$) which does not reflect the complex density profile and multiphase structure seen in the CGM. This results in the outflow interacting with less mass once it leaves the disc, and thus producing only weak X-ray emission. Additionally, the CGM is expected to contain in/outflowing clouds \citep{DiTeodoro2020,Nelson2020,Rahul2023} which could also induce mixing-driven X-ray emission as the outflow interacts with them. This possible interaction pathway is currently missed in our static CGM modelling.

However, our clumpy ISM setup allows the wind to vent through low density regions, resulting in less ejection of gas into the CGM than in a homogeneous or unresolved ISM setup. This could suggest that it is in fact challenging to inflate large X-ray bubbles in the halos of galaxies. Furthermore, some previous studies that have observed large X-ray bubbles in the CGM \citep{Lansbury2018} have also found spatially coincident radio emission consistent with a jet \citep{Harrison2015}. Additionally, direct radiation pressure from the AGN could be responsible for driving large outflowing bubbles \citep{Somalwar2020}. These studies could suggest that additional energy injection is required to create a bright X-ray bubble. However, in another system with a bright bubble \citep{Greene2014} found that the outflow could be described with a wind alone. 

Using a similar idealised clumpy ISM setup to our work, \cite{Tanner2025} investigated the extra-planar outflow from both a starburst and an AGN-driven jet, as well as a combination of both. They found that the inclusion of a starburst-driven mechanism was more effective at lifting material from the disc into a CGM-scale outflow than an AGN-only driver. They also investigated the predicted X-ray emission from such extra-planar outflows and, even in their combined AGN and starburst case, found that \textit{Chandra} struggled to observe CGM-scale bubbles even in a nearby ($d\simeq 4.3 \rm{\ Mpc}$) galaxy, although, they found that \textit{NewAthena} or \textit{AXIS} would be able to observe such kpc-scale outflows.

Future simulations with a more realistic treatment of both the ISM and the CGM are therefore vital to explore whether energy injection from winds alone is sufficient to inflate luminous X-ray bubbles.

\subsubsection{Photoionisation} \label{sec:photoionisation}

In this study, we have assumed collisional ionisation equilibrium (CIE) for the X-ray emitting gas and neglected to model photoionisation, with the outflow-driven shocks directly causing all the X-ray emission characterised in this work. However, photoionisation from the AGN is also expected to be an important contributor to the X-ray emission in active galaxies. This is evidenced by soft X-ray structures often being aligned with optical emission line ionisation cones \citep[e.g.,]{Bianchi2006,Paggi2012,WangChen2024} with emission line diagnostics in these regions (such as [\ion{O}{III}]/H$\beta$) suggesting photoionisation as the dominant X-ray emission mechanism \citep[see review in][]{Fabbiano2022chapter}.

However, other studies have found that the observed kiloparsec-scale X-ray emission can be described by either a photo- or collisional-ionisation spectral model \citep{Wang2010,DiGesu2017}. Other studies note that neither model alone can provide an adequate fit \citep[e.g.,][]{Travascio2021}, but spectral components from both models are needed; for example, \cite{Fabbiano2022} could not distinguish between photoionised and shock-heated origins for the extended soft X-ray emission observed in NGC 1167, but noted that the presence of high-excitation neon (\ion{Ne}{IX-X}) lines suggested at least some ionisation due to shocks. \cite{TrindadeFalcao2025} fit their extended emission with both a photoionisation and CIE model and found that the soft X-rays were dominated by collisional contributions. 
This diversity of observational results indicates that the dominant X-ray emission mechanism can differ, not only between galaxies, but also across regions within a single system \citep[e.g.,][]{Fabbiano2018}.

To break this degeneracy between the models and unlock understanding of the conditions in which collisional or photoionisation would dominate, more detailed theoretical predictions are required. One approach would be to include radiative transfer modelling to track the spectra of the photoionised plasma, alongside the collisionally ionised gas from the outflow. Using a similar idealised setup to ours, \cite{Meenakshi2022a,Meenakshi2022b} post-processed their jet-ISM simulations with \textsc{Cloudy} photoionisation models. They found that the fraction of dense ($n>100\ \rm{cm^{-3}}$) gas affected by photoionisation was much lower (only $\approx 5\%$) than the gas that was collisionally ionised by their jet-induced shock, suggesting that, at least for our similarly idealised, clumpy setup, collisional emission may be the dominant mechanism.

Alongside collisional and photoionisation, there are additional effects that are less well-understood but which may contribute to the observed X-ray emission on extended scales. X-ray emission from the AGN itself may also scatter more widely, an effect confirmed by UV polarimetry and spectroscopy \citep[e.g.,][]{Tadhunter2002}. Additionally, non-equilibrium cooling effects are likely to play a role in galactic outflows: for example, \cite{Sarkar2022} found that their simulated, self-ionising star formation-driven galactic outflow had enhanced X-ray emission compared to equilibrium models. Although more commonly discussed in the context of starburst-driven winds \citep[e.g.][]{Liu2012,Zhang2014}, charge exchange (CX) may also occur in AGN outflows, tracing interfaces between the ionised and neutral phases. Charge exchange has been proposed as a contributor to the observed soft X-ray emission in several AGN-driven systems \citep[e.g.][]{Yang2020,Lopez2025}, potentially accounting for up to $\sim30\%$ of the total soft X-ray emission in those cases.

More theoretical work is therefore needed to fully explore the effect of photoionisation and non-equilibrium effects in AGN wind-driven systems. Simulations that focuses on jointly modelling shock-driven and photoionised spectral features, with a focus on lines that can distinguish between the two cases such as \ion{Ne}{IX-X}, \ion{Fe}{XXV}, and [\ion{O}{III}]/H$\beta$, would be valuable for the interpretation of observed X-ray emission in AGN.

\section{Conclusions and Outlook} \label{sec:conc}

In this study, we used simulations of an AGN wind interacting with a clumpy ISM  \citep{Ward2024} and investigated the resulting collisional X-ray emission, considering both Bremsstrahlung and metal line contributions. Our main results are as follows:

\begin{itemize}
    \item Wind-ISM interactions produce luminous X-ray emission due to mixing between the AGN wind and ISM gas. This X-ray emission is strongest in the `tails' behind dense clumps, where the stripped gas from the clouds is efficiently mixed with the wind. This is in contrast to analytic results for a wind in a homogeneous medium which find that the emission is dominated by shocked ISM gas.
    \item The inclusion of metal line cooling results in slightly larger outflowing cold clouds and boosts the X-ray contribution of the mixed gas phase, thanks to strong line emission in the soft band.
    \item We predict a strong scaling between the AGN luminosity and the resulting extended X-ray emission, with $L_X \propto L_{\rm{AGN}}$. This demonstrates our outflow is in the rapid cooling limit where the shocked ambient medium cools on the outflow timescale \citep{Nims2015}.
    \item This emission could be detectable in nearby galaxies as spatially extended X-rays from the central 1-4 kpc. The emission from wind-ISM interactions can be detected over the diffuse contribution from star formation, but is dominated by the central point source X-ray emission from the AGN itself. This emission could be resolved up to a distance of $d_{\rm{A}}\lesssim 410\ \rm{Mpc}$ ($z\lesssim 0.11$) for an instrument with a resolution of $0.5"$. However, it will be challenging to separate this emission from that caused by photoionisation Future stimulations that can jointly model the spectral signatures of collisional and photoionisation will be critical in interpreting the origin of extended X-ray emission in AGN outflows.
\end{itemize}

This study represents a first look at the X-ray emission expected from wind-cloud interactions, and a proof-of-concept for being able to observe this emission with high spatial resolution X-ray observatories such as \textit{Chandra}. Looking ahead, the proposed observatories \textit{AXIS}\footnote{\url{https://axis.umd.edu/}} and \textit{Lynx}\footnote{\url{https://www.lynxobservatory.com/}} will also both have excellent spatial resolution and sensitivity in the soft band, making them ideal instruments for studying AGN-ISM interactions and galactic outflows.

The results of this work also have interesting implications for our theoretical understanding of AGN outflows. The existence of cold clouds surviving in galactic outflows has been a longstanding puzzle, as the cloud crushing timescale is expected to be shorter than the acceleration time \citep[e.g.,][]{Klein1994,Zhang2017}. Wind tunnel simulations have demonstrated that cold clouds can survive, and even grow their mass, under strong ram pressure thanks to radiative cooling at the mixing layer between the hot wind and cool cloud \citep[e.g.,][]{Gronke2018,BandaBarragan2020,Gronke2020,Mandal2024,Tan2024}. As shown in \cite{Ward2024}, our simulations are able to entrain cold clouds on $\geq 5\ \rm{Myr}$ timescales.  Further work in this area has been performed in Almeida et al. 2025 (sub. MNRAS), who add an enhanced refinement criteria to the cold clouds in \cite{Ward2024}, finding that the density of these outflowing clouds is significantly more sensitive to the AGN wind-driven mixing layers than to the initial ISM conditions. In the present work we demonstrated that X-ray emission is a powerful tracer in understanding the mixing, cooling and survival of cold clouds in a galactic wind, showing that the X-ray luminosity is dominated by the rapidly-cooling mixed wind-ISM phase.

Additionally, the inclusion of metal line cooling in our simulations enables a range of additional observational predictions. In particular, rest-frame UV \ion{O}{VI} is expected to be produced in radiative mixing layers \citep[e.g.,][]{Sembach2000,Kwak2010,Suoqing2019} such as our wind-ISM interactions. Optical [\ion{O}{III}] could be used to study the wind-driven shocks in our clumpy medium. Intriguingly, the existence of outflowing cold gas clouds in our own Milky Way \citep{DiTeodoro2018,DiTeodoro2020} could provide a local test-bed for these mechanisms. Our simulations predict that these individual clouds should exhibit X-ray emitting tails as they interact with the galactic wind. These features could be detectable by telescopes such as \textit{XMM-Newton} or the proposed \textit{NewAthena}\footnote{\url{www.the-athena-x-ray-observatory.eu}} mission, given their large effective areas and high soft-band sensitivity.

Together, these results offer a path forward for observationally testing wind-ISM mixing and cold cloud entrainment, which will help advance our understanding of multiphase outflows, both within the Milky Way and beyond.

\section*{Acknowledgements}

The authors would like to thank the referee, Dr Dipanjan Mukherjee, for a helpful and constructive report. We also thank Greg Bryan and Lachlan Lancaster for insightful discussions on metal line cooling and numerical convergence that helped to shape this work. 

CH acknowledges funding from an United Kingdom Research and Innovation grant (code: MR/V022830/1). This work used computing facilities from the Computational Center for Particle and Astrophysics (C2PAP), part of the ORIGINS cluster which is funded by the Deutsche Forschungsgemeinschaft (DFG; German Research Foundation) under Germany's Excellence Strategy: EXC-2094-390783311. This work also used the DiRAC Memory Intensive service (Cosma8) at Durham University, managed by the Institute for Computational Cosmology on behalf of the STFC DiRAC HPC Facility (www.dirac.ac.uk). The DiRAC service at Durham was funded by BEIS, UKRI and STFC capital funding, Durham University and STFC operations grants. DiRAC is part of the UKRI Digital Research Infrastructure.

In addition to those referenced in the text, this work also made use of the following software packages: \texttt{astropy} \citep{astropy:2013,astropy:2018,astropy:2022}, \texttt{matplotlib} \citep{Hunter:2007}, \texttt{numpy} \citep{numpy}, \texttt{python} \citep{python}, \texttt{scipy} \citep{2020SciPy-NMeth,scipy_17467817}, and \texttt{CMasher} \citep{2020JOSS....5.2004V,CMasher_14186007}. This research has made use of the Astrophysics Data System, funded by NASA under Cooperative Agreement 80NSSC21M00561. Software citation information aggregated using \texttt{\href{https://www.tomwagg.com/software-citation-station/}{The Software Citation Station}} \citep{software-citation-station-paper,software-citation-station-zenodo}.

\section*{Data Availability}
The data underlying this paper will be shared on reasonable request
to the corresponding author.


\bibliographystyle{mnras}
\bibliography{lib.bib}

\begin{thebibliography}{}
\makeatletter
\relax
\def\mn@urlcharsother{\let\do\@makeother \do\$\do\&\do\#\do\^\do\_\do\%\do\~}
\def\mn@doi{\begingroup\mn@urlcharsother \@ifnextchar [ {\mn@doi@} {\mn@doi@[]}}
\def\mn@doi@[#1]#2{\def\@tempa{#1}\ifx\@tempa\@empty \href {http://dx.doi.org/#2} {doi:#2}\else \href {http://dx.doi.org/#2} {#1}\fi \endgroup}
\def\mn@eprint#1#2{\mn@eprint@#1:#2::\@nil}
\def\mn@eprint@arXiv#1{\href {http://arxiv.org/abs/#1} {{\tt arXiv:#1}}}
\def\mn@eprint@dblp#1{\href {http://dblp.uni-trier.de/rec/bibtex/#1.xml} {dblp:#1}}
\def\mn@eprint@#1:#2:#3:#4\@nil{\def\@tempa {#1}\def\@tempb {#2}\def\@tempc {#3}\ifx \@tempc \@empty \let \@tempc \@tempb \let \@tempb \@tempa \fi \ifx \@tempb \@empty \def\@tempb {arXiv}\fi \@ifundefined {mn@eprint@\@tempb}{\@tempb:\@tempc}{\expandafter \expandafter \csname mn@eprint@\@tempb\endcsname \expandafter{\@tempc}}}

\bibitem[\protect\citeauthoryear{{Abruzzo}, {Fielding}  \& {Bryan}}{{Abruzzo} et~al.}{2024}]{Abruzzo2024}
{Abruzzo} M.~W.,  {Fielding} D.~B.,   {Bryan} G.~L.,  2024, \mn@doi [\apj] {10.3847/1538-4357/ad1e51}, \href {https://ui.adsabs.harvard.edu/abs/2024ApJ...966..181A} {966, 181}

\bibitem[\protect\citeauthoryear{{Astropy Collaboration} et~al.,}{{Astropy Collaboration} et~al.}{2013}]{astropy:2013}
{Astropy Collaboration} et~al., 2013, \mn@doi [\aap] {10.1051/0004-6361/201322068}, \href {http://adsabs.harvard.edu/abs/2013A%26A...558A..33A} {558, A33}

\bibitem[\protect\citeauthoryear{{Astropy Collaboration} et~al.,}{{Astropy Collaboration} et~al.}{2018}]{astropy:2018}
{Astropy Collaboration} et~al., 2018, \mn@doi [\aj] {10.3847/1538-3881/aabc4f}, \href {https://ui.adsabs.harvard.edu/abs/2018AJ....156..123A} {156, 123}

\bibitem[\protect\citeauthoryear{{Astropy Collaboration} et~al.,}{{Astropy Collaboration} et~al.}{2022}]{astropy:2022}
{Astropy Collaboration} et~al., 2022, \mn@doi [\apj] {10.3847/1538-4357/ac7c74}, \href {https://ui.adsabs.harvard.edu/abs/2022ApJ...935..167A} {935, 167}

\bibitem[\protect\citeauthoryear{{Audibert} et~al.,}{{Audibert} et~al.}{2023}]{Audibert2023}
{Audibert} A.,  et~al., 2023, \mn@doi [\aap] {10.1051/0004-6361/202345964}, \href {https://ui.adsabs.harvard.edu/abs/2023A&A...671L..12A} {671, L12}

\bibitem[\protect\citeauthoryear{{Banda-Barrag{\'a}n}, {Br{\"u}ggen}, {Federrath}, {Wagner}, {Scannapieco}  \& {Cottle}}{{Banda-Barrag{\'a}n} et~al.}{2020}]{BandaBarragan2020}
{Banda-Barrag{\'a}n} W.~E.,  {Br{\"u}ggen} M.,  {Federrath} C.,  {Wagner} A.~Y.,  {Scannapieco} E.,   {Cottle} J.,  2020, \mn@doi [\mnras] {10.1093/mnras/staa2904}, \href {https://ui.adsabs.harvard.edu/abs/2020MNRAS.499.2173B} {499, 2173}

\bibitem[\protect\citeauthoryear{{Banda-Barrag{\'a}n}, {Br{\"u}ggen}, {Heesen}, {Scannapieco}, {Cottle}, {Federrath}  \& {Wagner}}{{Banda-Barrag{\'a}n} et~al.}{2021}]{BandaBarragan2021}
{Banda-Barrag{\'a}n} W.~E.,  {Br{\"u}ggen} M.,  {Heesen} V.,  {Scannapieco} E.,  {Cottle} J.,  {Federrath} C.,   {Wagner} A.~Y.,  2021, \mn@doi [\mnras] {10.1093/mnras/stab1884}, \href {https://ui.adsabs.harvard.edu/abs/2021MNRAS.506.5658B} {506, 5658}

\bibitem[\protect\citeauthoryear{{Bennett} \& {Sijacki}}{{Bennett} \& {Sijacki}}{2022}]{Bennett2022}
{Bennett} J.~S.,  {Sijacki} D.,  2022, \mn@doi [\mnras] {10.1093/mnras/stac1216}, \href {https://ui.adsabs.harvard.edu/abs/2022MNRAS.514..313B} {514, 313}

\bibitem[\protect\citeauthoryear{{Bennett}, {Sijacki}, {Costa}, {Laporte}  \& {Witten}}{{Bennett} et~al.}{2024}]{Bennett2024}
{Bennett} J.~S.,  {Sijacki} D.,  {Costa} T.,  {Laporte} N.,   {Witten} C.,  2024, \mn@doi [\mnras] {10.1093/mnras/stad3179}, \href {https://ui.adsabs.harvard.edu/abs/2024MNRAS.527.1033B} {527, 1033}

\bibitem[\protect\citeauthoryear{{Bertola} et~al.,}{{Bertola} et~al.}{2025}]{Bertola2025}
{Bertola} E.,  et~al., 2025, \mn@doi [\aap] {10.1051/0004-6361/202554281}, \href {https://ui.adsabs.harvard.edu/abs/2025A&A...699A.220B} {699, A220}

\bibitem[\protect\citeauthoryear{{Bianchi}, {Guainazzi}  \& {Chiaberge}}{{Bianchi} et~al.}{2006}]{Bianchi2006}
{Bianchi} S.,  {Guainazzi} M.,   {Chiaberge} M.,  2006, \mn@doi [\aap] {10.1051/0004-6361:20054091}, \href {https://ui.adsabs.harvard.edu/abs/2006A&A...448..499B} {448, 499}

\bibitem[\protect\citeauthoryear{{Bieri}, {Dubois}, {Rosdahl}, {Wagner}, {Silk}  \& {Mamon}}{{Bieri} et~al.}{2017}]{Bieri2017}
{Bieri} R.,  {Dubois} Y.,  {Rosdahl} J.,  {Wagner} A.,  {Silk} J.,   {Mamon} G.~A.,  2017, \mn@doi [\mnras] {10.1093/mnras/stw2380}, \href {https://ui.adsabs.harvard.edu/abs/2017MNRAS.464.1854B} {464, 1854}

\bibitem[\protect\citeauthoryear{{Biffi}, {Dolag}, {B{\"o}hringer}  \& {Lemson}}{{Biffi} et~al.}{2012}]{Biffi2012}
{Biffi} V.,  {Dolag} K.,  {B{\"o}hringer} H.,   {Lemson} G.,  2012, \mn@doi [\mnras] {10.1111/j.1365-2966.2011.20278.x}, \href {https://ui.adsabs.harvard.edu/abs/2012MNRAS.420.3545B} {420, 3545}

\bibitem[\protect\citeauthoryear{{Biffi}, {Dolag}  \& {B{\"o}hringer}}{{Biffi} et~al.}{2013}]{Biffi2013}
{Biffi} V.,  {Dolag} K.,   {B{\"o}hringer} H.,  2013, \mn@doi [\mnras] {10.1093/mnras/sts120}, \href {https://ui.adsabs.harvard.edu/abs/2013MNRAS.428.1395B} {428, 1395}

\bibitem[\protect\citeauthoryear{{Bischetti} et~al.,}{{Bischetti} et~al.}{2019}]{Bischetti2019}
{Bischetti} M.,  et~al., 2019, \mn@doi [\aap] {10.1051/0004-6361/201935524}, \href {https://ui.adsabs.harvard.edu/abs/2019A&A...628A.118B} {628, A118}

\bibitem[\protect\citeauthoryear{{Bischetti} et~al.,}{{Bischetti} et~al.}{2024}]{Bischetti2024}
{Bischetti} M.,  et~al., 2024, \mn@doi [\apj] {10.3847/1538-4357/ad4a77}, \href {https://ui.adsabs.harvard.edu/abs/2024ApJ...970....9B} {970, 9}

\bibitem[\protect\citeauthoryear{{Bower}, {Benson}, {Malbon}, {Helly}, {Frenk}, {Baugh}, {Cole}  \& {Lacey}}{{Bower} et~al.}{2006}]{Bower2006}
{Bower} R.~G.,  {Benson} A.~J.,  {Malbon} R.,  {Helly} J.~C.,  {Frenk} C.~S.,  {Baugh} C.~M.,  {Cole} S.,   {Lacey} C.~G.,  2006, \mn@doi [\mnras] {10.1111/j.1365-2966.2006.10519.x}, \href {https://ui.adsabs.harvard.edu/abs/2006MNRAS.370..645B} {370, 645}

\bibitem[\protect\citeauthoryear{{Churazov}, {Sazonov}, {Sunyaev}, {Forman}, {Jones}  \& {B{\"o}hringer}}{{Churazov} et~al.}{2005}]{Churazov2005}
{Churazov} E.,  {Sazonov} S.,  {Sunyaev} R.,  {Forman} W.,  {Jones} C.,   {B{\"o}hringer} H.,  2005, \mn@doi [\mnras] {10.1111/j.1745-3933.2005.00093.x}, \href {https://ui.adsabs.harvard.edu/abs/2005MNRAS.363L..91C} {363, L91}

\bibitem[\protect\citeauthoryear{{Cicone}, {Brusa}, {Ramos Almeida}, {Cresci}, {Husemann}  \& {Mainieri}}{{Cicone} et~al.}{2018}]{Cicone2018}
{Cicone} C.,  {Brusa} M.,  {Ramos Almeida} C.,  {Cresci} G.,  {Husemann} B.,   {Mainieri} V.,  2018, \mn@doi [Nature Astronomy] {10.1038/s41550-018-0406-3}, \href {https://ui.adsabs.harvard.edu/abs/2018NatAs...2..176C} {2, 176}

\bibitem[\protect\citeauthoryear{{Cooper}, {Bicknell}, {Sutherland}  \& {Bland-Hawthorn}}{{Cooper} et~al.}{2008}]{Cooper2008}
{Cooper} J.~L.,  {Bicknell} G.~V.,  {Sutherland} R.~S.,   {Bland-Hawthorn} J.,  2008, \mn@doi [\apj] {10.1086/524918}, \href {https://ui.adsabs.harvard.edu/abs/2008ApJ...674..157C} {674, 157}

\bibitem[\protect\citeauthoryear{{Costa}, {Sijacki}, {Trenti}  \& {Haehnelt}}{{Costa} et~al.}{2014a}]{Costa2014a}
{Costa} T.,  {Sijacki} D.,  {Trenti} M.,   {Haehnelt} M.~G.,  2014a, \mn@doi [\mnras] {10.1093/mnras/stu101}, \href {https://ui.adsabs.harvard.edu/abs/2014MNRAS.439.2146C} {439, 2146}

\bibitem[\protect\citeauthoryear{{Costa}, {Sijacki}  \& {Haehnelt}}{{Costa} et~al.}{2014b}]{Costa2014b}
{Costa} T.,  {Sijacki} D.,   {Haehnelt} M.~G.,  2014b, \mn@doi [\mnras] {10.1093/mnras/stu1632}, \href {https://ui.adsabs.harvard.edu/abs/2014MNRAS.444.2355C} {444, 2355}

\bibitem[\protect\citeauthoryear{Costa, Pakmor  \& Springel}{Costa et~al.}{2020}]{Costa2020}
Costa T.,  Pakmor R.,   Springel V.,  2020, \mn@doi [\mnras] {10.1093/mnras/staa2321}, 497, 5229

\bibitem[\protect\citeauthoryear{{Croston}, {Hardcastle}, {Kharb}, {Kraft}  \& {Hota}}{{Croston} et~al.}{2008}]{Croston2008}
{Croston} J.~H.,  {Hardcastle} M.~J.,  {Kharb} P.,  {Kraft} R.~P.,   {Hota} A.,  2008, \mn@doi [\apj] {10.1086/592268}, \href {https://ui.adsabs.harvard.edu/abs/2008ApJ...688..190C} {688, 190}

\bibitem[\protect\citeauthoryear{Davé, Anglés-Alcázar, Narayanan, Li, Rafieferantsoa  \& Appleby}{Davé et~al.}{2019}]{Dave2019}
Davé R.,  Anglés-Alcázar D.,  Narayanan D.,  Li Q.,  Rafieferantsoa M.~H.,   Appleby S.,  2019, \mn@doi [MNRAS] {10.1093/mnras/stz937}, 486, 2827

\bibitem[\protect\citeauthoryear{{Di Gesu}, {Costantini}, {Piconcelli}, {Kaastra}, {Mehdipour}  \& {Paltani}}{{Di Gesu} et~al.}{2017}]{DiGesu2017}
{Di Gesu} L.,  {Costantini} E.,  {Piconcelli} E.,  {Kaastra} J.~S.,  {Mehdipour} M.,   {Paltani} S.,  2017, \mn@doi [\aap] {10.1051/0004-6361/201731853}, \href {https://ui.adsabs.harvard.edu/abs/2017A&A...608A.115D} {608, A115}

\bibitem[\protect\citeauthoryear{{Di Teodoro}, {McClure-Griffiths}, {Lockman}, {Denbo}, {Endsley}, {Ford}  \& {Harrington}}{{Di Teodoro} et~al.}{2018}]{DiTeodoro2018}
{Di Teodoro} E.~M.,  {McClure-Griffiths} N.~M.,  {Lockman} F.~J.,  {Denbo} S.~R.,  {Endsley} R.,  {Ford} H.~A.,   {Harrington} K.,  2018, \mn@doi [\apj] {10.3847/1538-4357/aaad6a}, \href {https://ui.adsabs.harvard.edu/abs/2018ApJ...855...33D} {855, 33}

\bibitem[\protect\citeauthoryear{{Di Teodoro}, {McClure-Griffiths}, {Lockman}  \& {Armillotta}}{{Di Teodoro} et~al.}{2020}]{DiTeodoro2020}
{Di Teodoro} E.~M.,  {McClure-Griffiths} N.~M.,  {Lockman} F.~J.,   {Armillotta} L.,  2020, \mn@doi [\nat] {10.1038/s41586-020-2595-z}, \href {https://ui.adsabs.harvard.edu/abs/2020Natur.584..364D} {584, 364}

\bibitem[\protect\citeauthoryear{{Dolag} et~al.,}{{Dolag} et~al.}{2025}]{Dolag2025}
{Dolag} K.,  et~al., 2025, \mn@doi [arXiv e-prints] {10.48550/arXiv.2504.01061}, \href {https://ui.adsabs.harvard.edu/abs/2025arXiv250401061D} {p. arXiv:2504.01061}

\bibitem[\protect\citeauthoryear{{Dubois}, {Peirani}, {Pichon}, {Devriendt}, {Gavazzi}, {Welker}  \& {Volonteri}}{{Dubois} et~al.}{2016}]{Dubois2016}
{Dubois} Y.,  {Peirani} S.,  {Pichon} C.,  {Devriendt} J.,  {Gavazzi} R.,  {Welker} C.,   {Volonteri} M.,  2016, \mn@doi [\mnras] {10.1093/mnras/stw2265}, \href {https://ui.adsabs.harvard.edu/abs/2016MNRAS.463.3948D} {463, 3948}

\bibitem[\protect\citeauthoryear{{Dubois} et~al.,}{{Dubois} et~al.}{2021}]{Dubois2021}
{Dubois} Y.,  et~al., 2021, \mn@doi [\aap] {10.1051/0004-6361/202039429}, \href {https://ui.adsabs.harvard.edu/abs/2021A&A...651A.109D} {651, A109}

\bibitem[\protect\citeauthoryear{{Fabbiano} \& {Elvis}}{{Fabbiano} \& {Elvis}}{2022}]{Fabbiano2022chapter}
{Fabbiano} G.,  {Elvis} M.,  2022, in {Bambi} C.,  {Sangangelo} A.,  eds, , Handbook of X-ray and Gamma-ray Astrophysics.
p.~92, \mn@doi{10.1007/978-981-16-4544-0_111-1}

\bibitem[\protect\citeauthoryear{{Fabbiano}, {Paggi}, {Karovska}, {Elvis}, {Maksym}  \& {Wang}}{{Fabbiano} et~al.}{2018}]{Fabbiano2018}
{Fabbiano} G.,  {Paggi} A.,  {Karovska} M.,  {Elvis} M.,  {Maksym} W.~P.,   {Wang} J.,  2018, \mn@doi [\apj] {10.3847/1538-4357/aadc5d}, \href {https://ui.adsabs.harvard.edu/abs/2018ApJ...865...83F} {865, 83}

\bibitem[\protect\citeauthoryear{{Fabbiano} et~al.,}{{Fabbiano} et~al.}{2022}]{Fabbiano2022}
{Fabbiano} G.,  et~al., 2022, \mn@doi [\apj] {10.3847/1538-4357/ac8ff8}, \href {https://ui.adsabs.harvard.edu/abs/2022ApJ...938..105F} {938, 105}

\bibitem[\protect\citeauthoryear{Fabian}{Fabian}{2012}]{Fabian2012}
Fabian A.~C.,  2012, \mn@doi [\araa] {10.1146/annurev-astro-081811-125521}, 50, 455

\bibitem[\protect\citeauthoryear{{Faucher-Gigu{\`e}re} \& {Quataert}}{{Faucher-Gigu{\`e}re} \& {Quataert}}{2012}]{Faucher-Giguere2012}
{Faucher-Gigu{\`e}re} C.-A.,  {Quataert} E.,  2012, \mn@doi [\mnras] {10.1111/j.1365-2966.2012.21512.x}, \href {https://ui.adsabs.harvard.edu/abs/2012MNRAS.425..605F} {425, 605}

\bibitem[\protect\citeauthoryear{{Feruglio} et~al.,}{{Feruglio} et~al.}{2013}]{Feruglio2013}
{Feruglio} C.,  et~al., 2013, \mn@doi [\aap] {10.1051/0004-6361/201219746}, \href {https://ui.adsabs.harvard.edu/abs/2013A&A...549A..51F} {549, A51}

\bibitem[\protect\citeauthoryear{{Fielding}, {Ostriker}, {Bryan}  \& {Jermyn}}{{Fielding} et~al.}{2020}]{Fielding2020}
{Fielding} D.~B.,  {Ostriker} E.~C.,  {Bryan} G.~L.,   {Jermyn} A.~S.,  2020, \mn@doi [\apjl] {10.3847/2041-8213/ab8d2c}, \href {https://ui.adsabs.harvard.edu/abs/2020ApJ...894L..24F} {894, L24}

\bibitem[\protect\citeauthoryear{{Fluetsch} et~al.,}{{Fluetsch} et~al.}{2019}]{Fluetsch2019}
{Fluetsch} A.,  et~al., 2019, \mn@doi [\mnras] {10.1093/mnras/sty3449}, \href {https://ui.adsabs.harvard.edu/abs/2019MNRAS.483.4586F} {483, 4586}

\bibitem[\protect\citeauthoryear{{F{\"o}rster Schreiber} et~al.,}{{F{\"o}rster Schreiber} et~al.}{2019}]{ForsterSchreiber2019}
{F{\"o}rster Schreiber} N.~M.,  et~al., 2019, \mn@doi [\apj] {10.3847/1538-4357/ab0ca2}, \href {https://ui.adsabs.harvard.edu/abs/2019ApJ...875...21F} {875, 21}

\bibitem[\protect\citeauthoryear{{Girdhar} et~al.,}{{Girdhar} et~al.}{2022}]{Girdhar2022}
{Girdhar} A.,  et~al., 2022, \mn@doi [\mnras] {10.1093/mnras/stac073}, \href {https://ui.adsabs.harvard.edu/abs/2022MNRAS.tmp..170G} {}

\bibitem[\protect\citeauthoryear{{Gofford}, {Reeves}, {McLaughlin}, {Braito}, {Turner}, {Tombesi}  \& {Cappi}}{{Gofford} et~al.}{2015}]{Gofford2015}
{Gofford} J.,  {Reeves} J.~N.,  {McLaughlin} D.~E.,  {Braito} V.,  {Turner} T.~J.,  {Tombesi} F.,   {Cappi} M.,  2015, \mn@doi [\mnras] {10.1093/mnras/stv1207}, \href {https://ui.adsabs.harvard.edu/abs/2015MNRAS.451.4169G} {451, 4169}

\bibitem[\protect\citeauthoryear{Gommers et~al.,}{Gommers et~al.}{2025}]{scipy_17467817}
Gommers R.,  et~al., 2025, scipy/scipy: SciPy 1.16.3, \mn@doi{10.5281/zenodo.17467817}, \url {https://doi.org/10.5281/zenodo.17467817}

\bibitem[\protect\citeauthoryear{{Gonz{\'a}lez-Alfonso} et~al.,}{{Gonz{\'a}lez-Alfonso} et~al.}{2017}]{Gonzalez-Alfonso2017}
{Gonz{\'a}lez-Alfonso} E.,  et~al., 2017, \mn@doi [\apj] {10.3847/1538-4357/836/1/11}, \href {https://ui.adsabs.harvard.edu/abs/2017ApJ...836...11G} {836, 11}

\bibitem[\protect\citeauthoryear{{Greene}, {Pooley}, {Zakamska}, {Comerford}  \& {Sun}}{{Greene} et~al.}{2014}]{Greene2014}
{Greene} J.~E.,  {Pooley} D.,  {Zakamska} N.~L.,  {Comerford} J.~M.,   {Sun} A.-L.,  2014, \mn@doi [\apj] {10.1088/0004-637X/788/1/54}, \href {https://ui.adsabs.harvard.edu/abs/2014ApJ...788...54G} {788, 54}

\bibitem[\protect\citeauthoryear{{Gronke} \& {Oh}}{{Gronke} \& {Oh}}{2018}]{Gronke2018}
{Gronke} M.,  {Oh} S.~P.,  2018, \mn@doi [\mnras] {10.1093/mnrasl/sly131}, \href {https://ui.adsabs.harvard.edu/abs/2018MNRAS.480L.111G} {480, L111}

\bibitem[\protect\citeauthoryear{{Gronke} \& {Oh}}{{Gronke} \& {Oh}}{2020}]{Gronke2020}
{Gronke} M.,  {Oh} S.~P.,  2020, \mn@doi [\mnras] {10.1093/mnras/stz3332}, \href {https://ui.adsabs.harvard.edu/abs/2020MNRAS.492.1970G} {492, 1970}

\bibitem[\protect\citeauthoryear{{Gu} et~al.,}{{Gu} et~al.}{2019}]{Gu2019}
{Gu} L.,  et~al., 2019, \mn@doi [\aap] {10.1051/0004-6361/201833860}, \href {https://ui.adsabs.harvard.edu/abs/2019A&A...627A..51G} {627, A51}

\bibitem[\protect\citeauthoryear{Harris et~al.,}{Harris et~al.}{2020}]{numpy}
Harris C.~R.,  et~al., 2020, \mn@doi [Nature] {10.1038/s41586-020-2649-2}, 585, 357

\bibitem[\protect\citeauthoryear{Harrison \& Ramos~Almeida}{Harrison \& Ramos~Almeida}{2024}]{Harrison2024}
Harrison C.~M.,  Ramos~Almeida C.,  2024, \mn@doi [Galaxies] {10.3390/galaxies12020017}, 12

\bibitem[\protect\citeauthoryear{{Harrison}, {Thomson}, {Alexander}, {Bauer}, {Edge}, {Hogan}, {Mullaney}  \& {Swinbank}}{{Harrison} et~al.}{2015}]{Harrison2015}
{Harrison} C.~M.,  {Thomson} A.~P.,  {Alexander} D.~M.,  {Bauer} F.~E.,  {Edge} A.~C.,  {Hogan} M.~T.,  {Mullaney} J.~R.,   {Swinbank} A.~M.,  2015, \mn@doi [\apj] {10.1088/0004-637X/800/1/45}, \href {https://ui.adsabs.harvard.edu/abs/2015ApJ...800...45H} {800, 45}

\bibitem[\protect\citeauthoryear{Harrison, Costa, Tadhunter, Flütsch, Kakkad, Perna  \& Vietri}{Harrison et~al.}{2018}]{Harrison2018}
Harrison C.~M.,  Costa T.,  Tadhunter C.~N.,  Flütsch A.,  Kakkad D.,  Perna M.,   Vietri G.,  2018, \mn@doi [Nature Astronomy] {10.1038/s41550-018-0403-6}, 2, 198

\bibitem[\protect\citeauthoryear{Hunter}{Hunter}{2007}]{Hunter:2007}
Hunter J.~D.,  2007, \mn@doi [Computing in Science \& Engineering] {10.1109/MCSE.2007.55}, 9, 90

\bibitem[\protect\citeauthoryear{Jarvis et~al.,}{Jarvis et~al.}{2021}]{Jarvis2021}
Jarvis M.~E.,  et~al., 2021, \mn@doi [MNRAS] {10.1093/mnras/stab549}, 503, 1780

\bibitem[\protect\citeauthoryear{{Jennings}, {Babul}, {Dav{\'e}}, {Cui}  \& {Rennehan}}{{Jennings} et~al.}{2025}]{Jennings2025}
{Jennings} F.~J.,  {Babul} A.,  {Dav{\'e}} R.,  {Cui} W.,   {Rennehan} D.,  2025, \mn@doi [\mnras] {10.1093/mnras/stae2592}, \href {https://ui.adsabs.harvard.edu/abs/2025MNRAS.536..145J} {536, 145}

\bibitem[\protect\citeauthoryear{{Ji}, {Oh}  \& {Masterson}}{{Ji} et~al.}{2019}]{Suoqing2019}
{Ji} S.,  {Oh} S.~P.,   {Masterson} P.,  2019, \mn@doi [\mnras] {10.1093/mnras/stz1248}, \href {https://ui.adsabs.harvard.edu/abs/2019MNRAS.487..737J} {487, 737}

\bibitem[\protect\citeauthoryear{{King}}{{King}}{2003}]{King2003}
{King} A.,  2003, \mn@doi [\apjl] {10.1086/379143}, \href {https://ui.adsabs.harvard.edu/abs/2003ApJ...596L..27K} {596, L27}

\bibitem[\protect\citeauthoryear{{King}}{{King}}{2005}]{King2005}
{King} A.,  2005, \mn@doi [\apjl] {10.1086/499430}, \href {https://ui.adsabs.harvard.edu/abs/2005ApJ...635L.121K} {635, L121}

\bibitem[\protect\citeauthoryear{{Klein}, {McKee}  \& {Colella}}{{Klein} et~al.}{1994}]{Klein1994}
{Klein} R.~I.,  {McKee} C.~F.,   {Colella} P.,  1994, \mn@doi [\apj] {10.1086/173554}, \href {https://ui.adsabs.harvard.edu/abs/1994ApJ...420..213K} {420, 213}

\bibitem[\protect\citeauthoryear{{Kwak} \& {Shelton}}{{Kwak} \& {Shelton}}{2010}]{Kwak2010}
{Kwak} K.,  {Shelton} R.~L.,  2010, \mn@doi [\apj] {10.1088/0004-637X/719/1/523}, \href {https://ui.adsabs.harvard.edu/abs/2010ApJ...719..523K} {719, 523}

\bibitem[\protect\citeauthoryear{{Lamperti} et~al.,}{{Lamperti} et~al.}{2022}]{Lamperti2022}
{Lamperti} I.,  et~al., 2022, \mn@doi [\aap] {10.1051/0004-6361/202244054}, \href {https://ui.adsabs.harvard.edu/abs/2022A&A...668A..45L} {668, A45}

\bibitem[\protect\citeauthoryear{{Lancaster}, {Ostriker}, {Kim}  \& {Kim}}{{Lancaster} et~al.}{2021}]{Lancaster2021}
{Lancaster} L.,  {Ostriker} E.~C.,  {Kim} J.-G.,   {Kim} C.-G.,  2021, \mn@doi [\apj] {10.3847/1538-4357/abf8ab}, \href {https://ui.adsabs.harvard.edu/abs/2021ApJ...914...89L} {914, 89}

\bibitem[\protect\citeauthoryear{{Lancaster}, {Ostriker}, {Kim}, {Kim}  \& {Bryan}}{{Lancaster} et~al.}{2024}]{Lancaster2024}
{Lancaster} L.,  {Ostriker} E.~C.,  {Kim} C.-G.,  {Kim} J.-G.,   {Bryan} G.~L.,  2024, \mn@doi [\apj] {10.3847/1538-4357/ad47f6}, \href {https://ui.adsabs.harvard.edu/abs/2024ApJ...970...18L} {970, 18}

\bibitem[\protect\citeauthoryear{{Lansbury}, {Jarvis}, {Harrison}, {Alexander}, {Del Moro}, {Edge}, {Mullaney}  \& {Thomson}}{{Lansbury} et~al.}{2018}]{Lansbury2018}
{Lansbury} G.~B.,  {Jarvis} M.~E.,  {Harrison} C.~M.,  {Alexander} D.~M.,  {Del Moro} A.,  {Edge} A.~C.,  {Mullaney} J.~R.,   {Thomson} A.~P.,  2018, \mn@doi [\apjl] {10.3847/2041-8213/aab357}, \href {https://ui.adsabs.harvard.edu/abs/2018ApJ...856L...1L} {856, L1}

\bibitem[\protect\citeauthoryear{{Lehmer}, {Alexander}, {Bauer}, {Brandt}, {Goulding}, {Jenkins}, {Ptak}  \& {Roberts}}{{Lehmer} et~al.}{2010}]{Lehmer2010}
{Lehmer} B.~D.,  {Alexander} D.~M.,  {Bauer} F.~E.,  {Brandt} W.~N.,  {Goulding} A.~D.,  {Jenkins} L.~P.,  {Ptak} A.,   {Roberts} T.~P.,  2010, \mn@doi [\apj] {10.1088/0004-637X/724/1/559}, \href {https://ui.adsabs.harvard.edu/abs/2010ApJ...724..559L} {724, 559}

\bibitem[\protect\citeauthoryear{{Lewis} \& {Austin}}{{Lewis} \& {Austin}}{2002}]{LewisAustin2002}
{Lewis} G.~M.,  {Austin} P.~H.,  2002, 11th Conference on Atmospheric Radiation, ed. G. H. Smith \& J. P. Brodie, American Meteorological Society Conference Series, 123

\bibitem[\protect\citeauthoryear{{Liu}, {Wang}  \& {Mao}}{{Liu} et~al.}{2012}]{Liu2012}
{Liu} J.,  {Wang} Q.~D.,   {Mao} S.,  2012, \mn@doi [\mnras] {10.1111/j.1365-2966.2011.20263.x}, \href {https://ui.adsabs.harvard.edu/abs/2012MNRAS.420.3389L} {420, 3389}

\bibitem[\protect\citeauthoryear{{Longinotti} et~al.,}{{Longinotti} et~al.}{2023}]{Longinotti2023}
{Longinotti} A.~L.,  et~al., 2023, \mn@doi [\mnras] {10.1093/mnras/stad540}, \href {https://ui.adsabs.harvard.edu/abs/2023MNRAS.521.2134L} {521, 2134}

\bibitem[\protect\citeauthoryear{{Lopez}, {Lopez}, {Lanz}, {Otter}  \& {Alatalo}}{{Lopez} et~al.}{2025}]{Lopez2025}
{Lopez} S.,  {Lopez} L.~A.,  {Lanz} L.,  {Otter} J.~A.,   {Alatalo} K.,  2025, \mn@doi [arXiv e-prints] {10.48550/arXiv.2510.01321}, \href {https://ui.adsabs.harvard.edu/abs/2025arXiv251001321L} {p. arXiv:2510.01321}

\bibitem[\protect\citeauthoryear{{Mandal}, {Mukherjee}, {Federrath}, {Bicknell}, {Nesvadba}  \& {Mignone}}{{Mandal} et~al.}{2024}]{Mandal2024}
{Mandal} A.,  {Mukherjee} D.,  {Federrath} C.,  {Bicknell} G.~V.,  {Nesvadba} N. P.~H.,   {Mignone} A.,  2024, \mn@doi [\mnras] {10.1093/mnras/stae1295}, \href {https://ui.adsabs.harvard.edu/abs/2024MNRAS.531.2079M} {531, 2079}

\bibitem[\protect\citeauthoryear{{Marasco} et~al.,}{{Marasco} et~al.}{2020}]{Marasco2020}
{Marasco} A.,  et~al., 2020, \mn@doi [\aap] {10.1051/0004-6361/202038889}, \href {https://ui.adsabs.harvard.edu/abs/2020A&A...644A..15M} {644, A15}

\bibitem[\protect\citeauthoryear{Marconi, Risaliti, Gilli, Hunt, Maiolino  \& Salvati}{Marconi et~al.}{2004}]{Marconi2004}
Marconi A.,  Risaliti G.,  Gilli R.,  Hunt L.~K.,  Maiolino R.,   Salvati M.,  2004, \mn@doi [\mnras] {10.1111/j.1365-2966.2004.07765.x}, 351, 169

\bibitem[\protect\citeauthoryear{{Matzeu} et~al.,}{{Matzeu} et~al.}{2023}]{Matzeu2023}
{Matzeu} G.~A.,  et~al., 2023, \mn@doi [\aap] {10.1051/0004-6361/202245036}, \href {https://ui.adsabs.harvard.edu/abs/2023A&A...670A.182M} {670, A182}

\bibitem[\protect\citeauthoryear{{Meenakshi}, {Mukherjee}, {Wagner}, {Nesvadba}, {Morganti}, {Janssen}  \& {Bicknell}}{{Meenakshi} et~al.}{2022a}]{Meenakshi2022a}
{Meenakshi} M.,  {Mukherjee} D.,  {Wagner} A.~Y.,  {Nesvadba} N. P.~H.,  {Morganti} R.,  {Janssen} R. M.~J.,   {Bicknell} G.~V.,  2022a, \mn@doi [\mnras] {10.1093/mnras/stac167}, \href {https://ui.adsabs.harvard.edu/abs/2022MNRAS.511.1622M} {511, 1622}

\bibitem[\protect\citeauthoryear{{Meenakshi} et~al.,}{{Meenakshi} et~al.}{2022b}]{Meenakshi2022b}
{Meenakshi} M.,  et~al., 2022b, \mn@doi [\mnras] {10.1093/mnras/stac2251}, \href {https://ui.adsabs.harvard.edu/abs/2022MNRAS.516..766M} {516, 766}

\bibitem[\protect\citeauthoryear{Mukherjee, Bicknell, Sutherland  \& Wagner}{Mukherjee et~al.}{2016}]{Mukherjee2016}
Mukherjee D.,  Bicknell G.~V.,  Sutherland R.,   Wagner A.,  2016, \mn@doi [\mnras] {10.1093/mnras/stw1368}, 461, 967

\bibitem[\protect\citeauthoryear{{Nelson} et~al.,}{{Nelson} et~al.}{2020}]{Nelson2020}
{Nelson} D.,  et~al., 2020, \mn@doi [\mnras] {10.1093/mnras/staa2419}, \href {https://ui.adsabs.harvard.edu/abs/2020MNRAS.498.2391N} {498, 2391}

\bibitem[\protect\citeauthoryear{{Nims}, {Quataert}  \& {Faucher-Gigu{\`e}re}}{{Nims} et~al.}{2015}]{Nims2015}
{Nims} J.,  {Quataert} E.,   {Faucher-Gigu{\`e}re} C.-A.,  2015, \mn@doi [\mnras] {10.1093/mnras/stu2648}, \href {https://ui.adsabs.harvard.edu/abs/2015MNRAS.447.3612N} {447, 3612}

\bibitem[\protect\citeauthoryear{{Njeri} et~al.,}{{Njeri} et~al.}{2025}]{Njeri2025}
{Njeri} A.,  et~al., 2025, \mn@doi [\mnras] {10.1093/mnras/staf020}, \href {https://ui.adsabs.harvard.edu/abs/2025MNRAS.537..705N} {537, 705}

\bibitem[\protect\citeauthoryear{{Paggi}, {Wang}, {Fabbiano}, {Elvis}  \& {Karovska}}{{Paggi} et~al.}{2012}]{Paggi2012}
{Paggi} A.,  {Wang} J.,  {Fabbiano} G.,  {Elvis} M.,   {Karovska} M.,  2012, \mn@doi [\apj] {10.1088/0004-637X/756/1/39}, \href {https://ui.adsabs.harvard.edu/abs/2012ApJ...756...39P} {756, 39}

\bibitem[\protect\citeauthoryear{{Pakmor}, {Springel}, {Bauer}, {Mocz}, {Munoz}, {Ohlmann}, {Schaal}  \& {Zhu}}{{Pakmor} et~al.}{2016}]{Pakmor2016}
{Pakmor} R.,  {Springel} V.,  {Bauer} A.,  {Mocz} P.,  {Munoz} D.~J.,  {Ohlmann} S.~T.,  {Schaal} K.,   {Zhu} C.,  2016, \mn@doi [\mnras] {10.1093/mnras/stv2380}, \href {https://ui.adsabs.harvard.edu/abs/2016MNRAS.455.1134P} {455, 1134}

\bibitem[\protect\citeauthoryear{{Pereira-Santaella} et~al.,}{{Pereira-Santaella} et~al.}{2011}]{PereiraSantaella2011}
{Pereira-Santaella} M.,  et~al., 2011, \mn@doi [\aap] {10.1051/0004-6361/201117420}, \href {https://ui.adsabs.harvard.edu/abs/2011A&A...535A..93P} {535, A93}

\bibitem[\protect\citeauthoryear{{Pillepich}, {Nelson}, {Truong}, {Weinberger}, {Martin-Navarro}, {Springel}, {Faber}  \& {Hernquist}}{{Pillepich} et~al.}{2021}]{Pillepich2021}
{Pillepich} A.,  {Nelson} D.,  {Truong} N.,  {Weinberger} R.,  {Martin-Navarro} I.,  {Springel} V.,  {Faber} S.~M.,   {Hernquist} L.,  2021, \mn@doi [\mnras] {10.1093/mnras/stab2779}, \href {https://ui.adsabs.harvard.edu/abs/2021MNRAS.508.4667P} {508, 4667}

\bibitem[\protect\citeauthoryear{{Piotrowska}, {Bluck}, {Maiolino}  \& {Peng}}{{Piotrowska} et~al.}{2022}]{Piotrowska2022}
{Piotrowska} J.~M.,  {Bluck} A. F.~L.,  {Maiolino} R.,   {Peng} Y.,  2022, \mn@doi [\mnras] {10.1093/mnras/stab3673}, \href {https://ui.adsabs.harvard.edu/abs/2022MNRAS.512.1052P} {512, 1052}

\bibitem[\protect\citeauthoryear{{Planck Collaboration XIII}}{{Planck Collaboration XIII}}{2016}]{Planck2016cite}
{Planck Collaboration XIII} 2016, \mn@doi [\aap] {10.1051/0004-6361/201525830}, \href {https://ui.adsabs.harvard.edu/abs/2016A&A...594A..13P} {594, A13}

\bibitem[\protect\citeauthoryear{{Polack}, {Revalski}, {Crenshaw}, {Fischer}, {Schmitt}, {Kraemer}, {Meena}  \& {Rafelski}}{{Polack} et~al.}{2024}]{Polack2024}
{Polack} G.~E.,  {Revalski} M.,  {Crenshaw} D.~M.,  {Fischer} T.~C.,  {Schmitt} H.~R.,  {Kraemer} S.~B.,  {Meena} B.,   {Rafelski} M.,  2024, \mn@doi [\apj] {10.3847/1538-4357/ad71c3}, \href {https://ui.adsabs.harvard.edu/abs/2024ApJ...975..129P} {975, 129}

\bibitem[\protect\citeauthoryear{{Ponti} et~al.,}{{Ponti} et~al.}{2019}]{Ponti2019}
{Ponti} G.,  et~al., 2019, \mn@doi [\nat] {10.1038/s41586-019-1009-6}, \href {https://ui.adsabs.harvard.edu/abs/2019Natur.567..347P} {567, 347}

\bibitem[\protect\citeauthoryear{{Ponti}, {Morris}, {Churazov}, {Heywood}  \& {Fender}}{{Ponti} et~al.}{2021}]{Ponti2021}
{Ponti} G.,  {Morris} M.~R.,  {Churazov} E.,  {Heywood} I.,   {Fender} R.~P.,  2021, \mn@doi [\aap] {10.1051/0004-6361/202039636}, \href {https://ui.adsabs.harvard.edu/abs/2021A&A...646A..66P} {646, A66}

\bibitem[\protect\citeauthoryear{{Predehl} et~al.,}{{Predehl} et~al.}{2020}]{Predehl2020}
{Predehl} P.,  et~al., 2020, \mn@doi [\nat] {10.1038/s41586-020-2979-0}, \href {https://ui.adsabs.harvard.edu/abs/2020Natur.588..227P} {588, 227}

\bibitem[\protect\citeauthoryear{{Ramesh}, {Nelson}  \& {Pillepich}}{{Ramesh} et~al.}{2023}]{Rahul2023}
{Ramesh} R.,  {Nelson} D.,   {Pillepich} A.,  2023, \mn@doi [\mnras] {10.1093/mnras/stad951}, \href {https://ui.adsabs.harvard.edu/abs/2023MNRAS.522.1535R} {522, 1535}

\bibitem[\protect\citeauthoryear{{Ramos Almeida}, {Piqueras L{\'o}pez}, {Villar-Mart{\'\i}n}  \& {Bessiere}}{{Ramos Almeida} et~al.}{2017}]{RamosAlmeida2017}
{Ramos Almeida} C.,  {Piqueras L{\'o}pez} J.,  {Villar-Mart{\'\i}n} M.,   {Bessiere} P.~S.,  2017, \mn@doi [\mnras] {10.1093/mnras/stx1287}, \href {https://ui.adsabs.harvard.edu/abs/2017MNRAS.470..964R} {470, 964}

\bibitem[\protect\citeauthoryear{{Ramos Almeida} et~al.,}{{Ramos Almeida} et~al.}{2022}]{RamosAlmeida2022}
{Ramos Almeida} C.,  et~al., 2022, \mn@doi [\aap] {10.1051/0004-6361/202141906}, \href {https://ui.adsabs.harvard.edu/abs/2022A&A...658A.155R} {658, A155}

\bibitem[\protect\citeauthoryear{{Rose}, {Tadhunter}, {Ramos Almeida}, {Rodr{\'\i}guez Zaur{\'\i}n}, {Santoro}  \& {Spence}}{{Rose} et~al.}{2018}]{Rose2018}
{Rose} M.,  {Tadhunter} C.,  {Ramos Almeida} C.,  {Rodr{\'\i}guez Zaur{\'\i}n} J.,  {Santoro} F.,   {Spence} R.,  2018, \mn@doi [\mnras] {10.1093/mnras/stx2590}, \href {https://ui.adsabs.harvard.edu/abs/2018MNRAS.474..128R} {474, 128}

\bibitem[\protect\citeauthoryear{{Sarkar}, {Sternberg}  \& {Gnat}}{{Sarkar} et~al.}{2022}]{Sarkar2022}
{Sarkar} K.~C.,  {Sternberg} A.,   {Gnat} O.,  2022, \mn@doi [\apj] {10.3847/1538-4357/ac9835}, \href {https://ui.adsabs.harvard.edu/abs/2022ApJ...940...44S} {940, 44}

\bibitem[\protect\citeauthoryear{{Scannapieco} \& {Oh}}{{Scannapieco} \& {Oh}}{2004}]{Scannapieco2004}
{Scannapieco} E.,  {Oh} S.~P.,  2004, \mn@doi [\apj] {10.1086/386542}, \href {https://ui.adsabs.harvard.edu/abs/2004ApJ...608...62S} {608, 62}

\bibitem[\protect\citeauthoryear{Schaye et~al.,}{Schaye et~al.}{2015}]{Schaye2015}
Schaye J.,  et~al., 2015, \mn@doi [\mnras] {10.1093/mnras/stu2058}, 446, 521

\bibitem[\protect\citeauthoryear{{Sembach} et~al.,}{{Sembach} et~al.}{2000}]{Sembach2000}
{Sembach} K.~R.,  et~al., 2000, \mn@doi [\apjl] {10.1086/312785}, \href {https://ui.adsabs.harvard.edu/abs/2000ApJ...538L..31S} {538, L31}

\bibitem[\protect\citeauthoryear{{Sijacki} \& {Springel}}{{Sijacki} \& {Springel}}{2006}]{Sijacki2006}
{Sijacki} D.,  {Springel} V.,  2006, \mn@doi [\mnras] {10.1111/j.1365-2966.2005.09860.x}, \href {https://ui.adsabs.harvard.edu/abs/2006MNRAS.366..397S} {366, 397}

\bibitem[\protect\citeauthoryear{{Somalwar}, {Johnson}, {Stern}, {Goulding}, {Greene}, {Zakamska}, {Alexandroff}  \& {Chen}}{{Somalwar} et~al.}{2020}]{Somalwar2020}
{Somalwar} J.,  {Johnson} S.~D.,  {Stern} J.,  {Goulding} A.~D.,  {Greene} J.~E.,  {Zakamska} N.~L.,  {Alexandroff} R.~M.,   {Chen} H.-W.,  2020, \mn@doi [\apjl] {10.3847/2041-8213/ab733d}, \href {https://ui.adsabs.harvard.edu/abs/2020ApJ...890L..28S} {890, L28}

\bibitem[\protect\citeauthoryear{{Somerville}, {Hopkins}, {Cox}, {Robertson}  \& {Hernquist}}{{Somerville} et~al.}{2008}]{Somerville2008}
{Somerville} R.~S.,  {Hopkins} P.~F.,  {Cox} T.~J.,  {Robertson} B.~E.,   {Hernquist} L.,  2008, \mn@doi [\mnras] {10.1111/j.1365-2966.2008.13805.x}, \href {https://ui.adsabs.harvard.edu/abs/2008MNRAS.391..481S} {391, 481}

\bibitem[\protect\citeauthoryear{{Speranza} et~al.,}{{Speranza} et~al.}{2024}]{Speranza2024}
{Speranza} G.,  et~al., 2024, \mn@doi [\aap] {10.1051/0004-6361/202347715}, \href {https://ui.adsabs.harvard.edu/abs/2024A&A...681A..63S} {681, A63}

\bibitem[\protect\citeauthoryear{{Springel}}{{Springel}}{2010}]{Springel2010}
{Springel} V.,  2010, \mn@doi [\mnras] {10.1111/j.1365-2966.2009.15715.x}, \href {https://ui.adsabs.harvard.edu/abs/2010MNRAS.401..791S} {401, 791}

\bibitem[\protect\citeauthoryear{Springel, Matteo  \& Hernquist}{Springel et~al.}{2005}]{Springel2005_BH}
Springel V.,  Matteo T.~D.,   Hernquist L.,  2005, \mn@doi [\mnras] {10.1111/j.1365-2966.2005.09238.x}, 361, 776

\bibitem[\protect\citeauthoryear{{Stanley} et~al.,}{{Stanley} et~al.}{2017}]{Stanley2017}
{Stanley} F.,  et~al., 2017, \mn@doi [\mnras] {10.1093/mnras/stx2121}, \href {https://ui.adsabs.harvard.edu/abs/2017MNRAS.472.2221S} {472, 2221}

\bibitem[\protect\citeauthoryear{{Sutherland} \& {Bicknell}}{{Sutherland} \& {Bicknell}}{2007}]{Sutherland2007}
{Sutherland} R.~S.,  {Bicknell} G.~V.,  2007, \mn@doi [\apjs] {10.1086/520640}, \href {https://ui.adsabs.harvard.edu/abs/2007ApJS..173...37S} {173, 37}

\bibitem[\protect\citeauthoryear{{Tadhunter}, {Dickson}, {Morganti}, {Robinson}, {Wills}, {Villar-Martin}  \& {Hughes}}{{Tadhunter} et~al.}{2002}]{Tadhunter2002}
{Tadhunter} C.,  {Dickson} R.,  {Morganti} R.,  {Robinson} T.~G.,  {Wills} K.,  {Villar-Martin} M.,   {Hughes} M.,  2002, \mn@doi [\mnras] {10.1046/j.1365-8711.2002.05153.x}, \href {https://ui.adsabs.harvard.edu/abs/2002MNRAS.330..977T} {330, 977}

\bibitem[\protect\citeauthoryear{{Tan} \& {Fielding}}{{Tan} \& {Fielding}}{2024}]{Tan2024}
{Tan} B.,  {Fielding} D.~B.,  2024, \mn@doi [\mnras] {10.1093/mnras/stad3793}, \href {https://ui.adsabs.harvard.edu/abs/2024MNRAS.527.9683T} {527, 9683}

\bibitem[\protect\citeauthoryear{{Tanner} \& {Weaver}}{{Tanner} \& {Weaver}}{2022}]{Tanner2022}
{Tanner} R.,  {Weaver} K.~A.,  2022, \mn@doi [\aj] {10.3847/1538-3881/ac4d23}, \href {https://ui.adsabs.harvard.edu/abs/2022AJ....163..134T} {163, 134}

\bibitem[\protect\citeauthoryear{{Tanner}, {Weaver}  \& {Ogorza{\l}ek}}{{Tanner} et~al.}{2025}]{Tanner2025}
{Tanner} R.,  {Weaver} K.~A.,   {Ogorza{\l}ek} A.,  2025, \mn@doi [\apj] {10.3847/1538-4357/ae00c4}, \href {https://ui.adsabs.harvard.edu/abs/2025ApJ...992...90T} {992, 90}

\bibitem[\protect\citeauthoryear{Terrazas, Bell, Woo  \& Henriques}{Terrazas et~al.}{2017}]{Terrazas2017}
Terrazas B.~A.,  Bell E.~F.,  Woo J.,   Henriques B. M.~B.,  2017, \mn@doi [\apj] {10.3847/1538-4357/aa7d07}, 844, 170

\bibitem[\protect\citeauthoryear{{Tombesi}, {Cappi}, {Reeves}, {Nemmen}, {Braito}, {Gaspari}  \& {Reynolds}}{{Tombesi} et~al.}{2013}]{Tombesi2013}
{Tombesi} F.,  {Cappi} M.,  {Reeves} J.~N.,  {Nemmen} R.~S.,  {Braito} V.,  {Gaspari} M.,   {Reynolds} C.~S.,  2013, \mn@doi [\mnras] {10.1093/mnras/sts692}, \href {https://ui.adsabs.harvard.edu/abs/2013MNRAS.430.1102T} {430, 1102}

\bibitem[\protect\citeauthoryear{{Travascio}, {Fabbiano}, {Paggi}, {Elvis}, {Maksym}, {Morganti}, {Oosterloo}  \& {Fiore}}{{Travascio} et~al.}{2021}]{Travascio2021}
{Travascio} A.,  {Fabbiano} G.,  {Paggi} A.,  {Elvis} M.,  {Maksym} W.~P.,  {Morganti} R.,  {Oosterloo} T.,   {Fiore} F.,  2021, \mn@doi [\apj] {10.3847/1538-4357/ac18c7}, \href {https://ui.adsabs.harvard.edu/abs/2021ApJ...921..129T} {921, 129}

\bibitem[\protect\citeauthoryear{{Trindade Falc{\~a}o}, {Middei}, {Fabbiano}, {Elvis}, {Zhu}, {Maksym}, {Kr{\'o}l}  \& {Feuillet}}{{Trindade Falc{\~a}o} et~al.}{2025}]{TrindadeFalcao2025}
{Trindade Falc{\~a}o} A.,  {Middei} R.,  {Fabbiano} G.,  {Elvis} M.,  {Zhu} P.,  {Maksym} W.~P.,  {Kr{\'o}l} D.~{\L}.,   {Feuillet} L.,  2025, \mn@doi [\apj] {10.3847/1538-4357/ae07d0}, \href {https://ui.adsabs.harvard.edu/abs/2025ApJ...993..247T} {993, 247}

\bibitem[\protect\citeauthoryear{Van~Rossum \& Drake}{Van~Rossum \& Drake}{2009}]{python}
Van~Rossum G.,  Drake F.~L.,  2009, Python 3 Reference Manual.
CreateSpace, Scotts Valley, CA

\bibitem[\protect\citeauthoryear{{Veilleux}, {Maiolino}, {Bolatto}  \& {Aalto}}{{Veilleux} et~al.}{2020}]{Veilleux2020}
{Veilleux} S.,  {Maiolino} R.,  {Bolatto} A.~D.,   {Aalto} S.,  2020, \mn@doi [\aapr] {10.1007/s00159-019-0121-9}, \href {https://ui.adsabs.harvard.edu/abs/2020A&ARv..28....2V} {28, 2}

\bibitem[\protect\citeauthoryear{Virtanen et~al.,}{Virtanen et~al.}{2020}]{2020SciPy-NMeth}
Virtanen P.,  et~al., 2020, \mn@doi [Nature Methods] {10.1038/s41592-019-0686-2}, \href {https://rdcu.be/b08Wh} {17, 261}

\bibitem[\protect\citeauthoryear{Vogelsberger, Genel, Sijacki, Torrey, Springel  \& Hernquist}{Vogelsberger et~al.}{2013}]{Vogelsberger2013}
Vogelsberger M.,  Genel S.,  Sijacki D.,  Torrey P.,  Springel V.,   Hernquist L.,  2013, \mn@doi [\mnras] {10.1093/mnras/stt1789}, 436, 3031

\bibitem[\protect\citeauthoryear{{Wagg} \& {Broekgaarden}}{{Wagg} \& {Broekgaarden}}{2024}]{software-citation-station-paper}
{Wagg} T.,  {Broekgaarden} F.~S.,  2024, arXiv e-prints, \href {https://ui.adsabs.harvard.edu/abs/2024arXiv240604405W} {p. arXiv:2406.04405}

\bibitem[\protect\citeauthoryear{Wagg, Broekgaarden  \& Gültekin}{Wagg et~al.}{2025}]{software-citation-station-zenodo}
Wagg T.,  Broekgaarden F.,   Gültekin K.,  2025, TomWagg/software-citation-station: v1.3, \mn@doi{10.5281/zenodo.17145205}, \url {https://doi.org/10.5281/zenodo.17145205}

\bibitem[\protect\citeauthoryear{Wagner \& Bicknell}{Wagner \& Bicknell}{2011}]{Wagner2011}
Wagner A.~Y.,  Bicknell G.~V.,  2011, \mn@doi [The Astrophysical Journal] {10.1088/0004-637X/728/1/29}, 728, 29

\bibitem[\protect\citeauthoryear{{Wagner}, {Bicknell}  \& {Umemura}}{{Wagner} et~al.}{2012}]{Wagner2012}
{Wagner} A.~Y.,  {Bicknell} G.~V.,   {Umemura} M.,  2012, \mn@doi [\apj] {10.1088/0004-637X/757/2/136}, \href {https://ui.adsabs.harvard.edu/abs/2012ApJ...757..136W} {757, 136}

\bibitem[\protect\citeauthoryear{{Wang}, {Fabbiano}, {Risaliti}, {Elvis}, {Mundell}, {Dumas}, {Schinnerer}  \& {Zezas}}{{Wang} et~al.}{2010}]{Wang2010}
{Wang} J.,  {Fabbiano} G.,  {Risaliti} G.,  {Elvis} M.,  {Mundell} C.~G.,  {Dumas} G.,  {Schinnerer} E.,   {Zezas} A.,  2010, \mn@doi [\apjl] {10.1088/2041-8205/719/2/L208}, \href {https://ui.adsabs.harvard.edu/abs/2010ApJ...719L.208W} {719, L208}

\bibitem[\protect\citeauthoryear{{Wang} et~al.,}{{Wang} et~al.}{2024a}]{Wang2024}
{Wang} W.,  et~al., 2024a, \mn@doi [\aap] {10.1051/0004-6361/202348531}, \href {https://ui.adsabs.harvard.edu/abs/2024A&A...683A.169W} {683, A169}

\bibitem[\protect\citeauthoryear{{Wang}, {Wang}, {Dadina}, {Fabbiano}, {Elvis}, {Bianchi}  \& {Guainazzi}}{{Wang} et~al.}{2024b}]{WangChen2024}
{Wang} C.,  {Wang} J.,  {Dadina} M.,  {Fabbiano} G.,  {Elvis} M.,  {Bianchi} S.,   {Guainazzi} M.,  2024b, \mn@doi [\apj] {10.3847/1538-4357/ad18c9}, \href {https://ui.adsabs.harvard.edu/abs/2024ApJ...962..188W} {962, 188}

\bibitem[\protect\citeauthoryear{{Ward}, {Harrison}, {Costa}  \& {Mainieri}}{{Ward} et~al.}{2022}]{Ward2022}
{Ward} S.~R.,  {Harrison} C.~M.,  {Costa} T.,   {Mainieri} V.,  2022, \mn@doi [\mnras] {10.1093/mnras/stac1219}, \href {https://ui.adsabs.harvard.edu/abs/2022MNRAS.514.2936W} {514, 2936}

\bibitem[\protect\citeauthoryear{{Ward}, {Costa}, {Harrison}  \& {Mainieri}}{{Ward} et~al.}{2024}]{Ward2024}
{Ward} S.~R.,  {Costa} T.,  {Harrison} C.~M.,   {Mainieri} V.,  2024, \mn@doi [\mnras] {10.1093/mnras/stae1816}, \href {https://ui.adsabs.harvard.edu/abs/2024MNRAS.533.1733W} {533, 1733}

\bibitem[\protect\citeauthoryear{Weinberger et~al.,}{Weinberger et~al.}{2018}]{Weinberger2018}
Weinberger R.,  et~al., 2018, \mn@doi [\mnras] {10.1093/mnras/sty1733}, 479, 4056

\bibitem[\protect\citeauthoryear{{Weinberger}, {Springel}  \& {Pakmor}}{{Weinberger} et~al.}{2020}]{Weinberger2020}
{Weinberger} R.,  {Springel} V.,   {Pakmor} R.,  2020, \mn@doi [\apjs] {10.3847/1538-4365/ab908c}, \href {https://ui.adsabs.harvard.edu/abs/2020ApJS..248...32W} {248, 32}

\bibitem[\protect\citeauthoryear{{Wellons} et~al.,}{{Wellons} et~al.}{2023}]{Wellons2023}
{Wellons} S.,  et~al., 2023, \mn@doi [\mnras] {10.1093/mnras/stad511}, \href {https://ui.adsabs.harvard.edu/abs/2023MNRAS.520.5394W} {520, 5394}

\bibitem[\protect\citeauthoryear{{Yang}, {Zhang}  \& {Ji}}{{Yang} et~al.}{2020}]{Yang2020}
{Yang} H.,  {Zhang} S.,   {Ji} L.,  2020, \mn@doi [\apj] {10.3847/1538-4357/ab80c9}, \href {https://ui.adsabs.harvard.edu/abs/2020ApJ...894...22Y} {894, 22}

\bibitem[\protect\citeauthoryear{{Zhang}, {Wang}, {Ji}, {Smith}, {Foster}  \& {Zhou}}{{Zhang} et~al.}{2014}]{Zhang2014}
{Zhang} S.,  {Wang} Q.~D.,  {Ji} L.,  {Smith} R.~K.,  {Foster} A.~R.,   {Zhou} X.,  2014, \mn@doi [\apj] {10.1088/0004-637X/794/1/61}, \href {https://ui.adsabs.harvard.edu/abs/2014ApJ...794...61Z} {794, 61}

\bibitem[\protect\citeauthoryear{{Zhang}, {Thompson}, {Quataert}  \& {Murray}}{{Zhang} et~al.}{2017}]{Zhang2017}
{Zhang} D.,  {Thompson} T.~A.,  {Quataert} E.,   {Murray} N.,  2017, \mn@doi [\mnras] {10.1093/mnras/stx822}, \href {https://ui.adsabs.harvard.edu/abs/2017MNRAS.468.4801Z} {468, 4801}

\bibitem[\protect\citeauthoryear{Zhuang, Ho  \& Shangguan}{Zhuang et~al.}{2021}]{Zhuang2021}
Zhuang M.-Y.,  Ho L.~C.,   Shangguan J.,  2021, \mn@doi [\apj] {10.3847/1538-4357/abc94d}, 906, 38

\bibitem[\protect\citeauthoryear{{ZuHone} \& {Hallman}}{{ZuHone} \& {Hallman}}{2016}]{ZuHone2016}
{ZuHone} J.~A.,  {Hallman} E.~J.,  2016, {pyXSIM: Synthetic X-ray observations generator}, Astrophysics Source Code Library, record ascl:1608.002

\bibitem[\protect\citeauthoryear{{ZuHone}, {Vikhlinin}, {Tremblay}, {Randall}, {Andrade-Santos}  \& {Bourdin}}{{ZuHone} et~al.}{2023}]{ZuHone2023}
{ZuHone} J.~A.,  {Vikhlinin} A.,  {Tremblay} G.~R.,  {Randall} S.~W.,  {Andrade-Santos} F.,   {Bourdin} H.,  2023, {SOXS: Simulated Observations of X-ray Sources}, Astrophysics Source Code Library, record ascl:2301.024

\bibitem[\protect\citeauthoryear{{Zubovas} \& {King}}{{Zubovas} \& {King}}{2012}]{Zubovas2012}
{Zubovas} K.,  {King} A.,  2012, \mn@doi [\apjl] {10.1088/2041-8205/745/2/L34}, \href {https://ui.adsabs.harvard.edu/abs/2012ApJ...745L..34Z} {745, L34}

\bibitem[\protect\citeauthoryear{{van der Velden}}{{van der Velden}}{2020}]{2020JOSS....5.2004V}
{van der Velden} E.,  2020, \mn@doi [The Journal of Open Source Software] {10.21105/joss.02004}, \href {https://ui.adsabs.harvard.edu/abs/2020JOSS....5.2004V} {5, 2004}

\bibitem[\protect\citeauthoryear{van~der Velden, Robert, Batten, Clauss, beskep, Ferdman, YANG  \& Thyng}{van~der Velden et~al.}{2024}]{CMasher_14186007}
van~der Velden E.,  Robert C.,  Batten A.,  Clauss C.,  beskep Ferdman E.,  YANG H.~D.,   Thyng K.,  2024, 1313e/CMasher: v1.9.2, \mn@doi{10.5281/zenodo.14186007}, \url {https://doi.org/10.5281/zenodo.14186007}

\makeatother
\end{thebibliography}


\appendix

\section{Metal line cooling and outflows}  \label{sec:ap:Zcooling}

\begin{figure}
    \centering
    \includegraphics[width=0.35\textwidth]{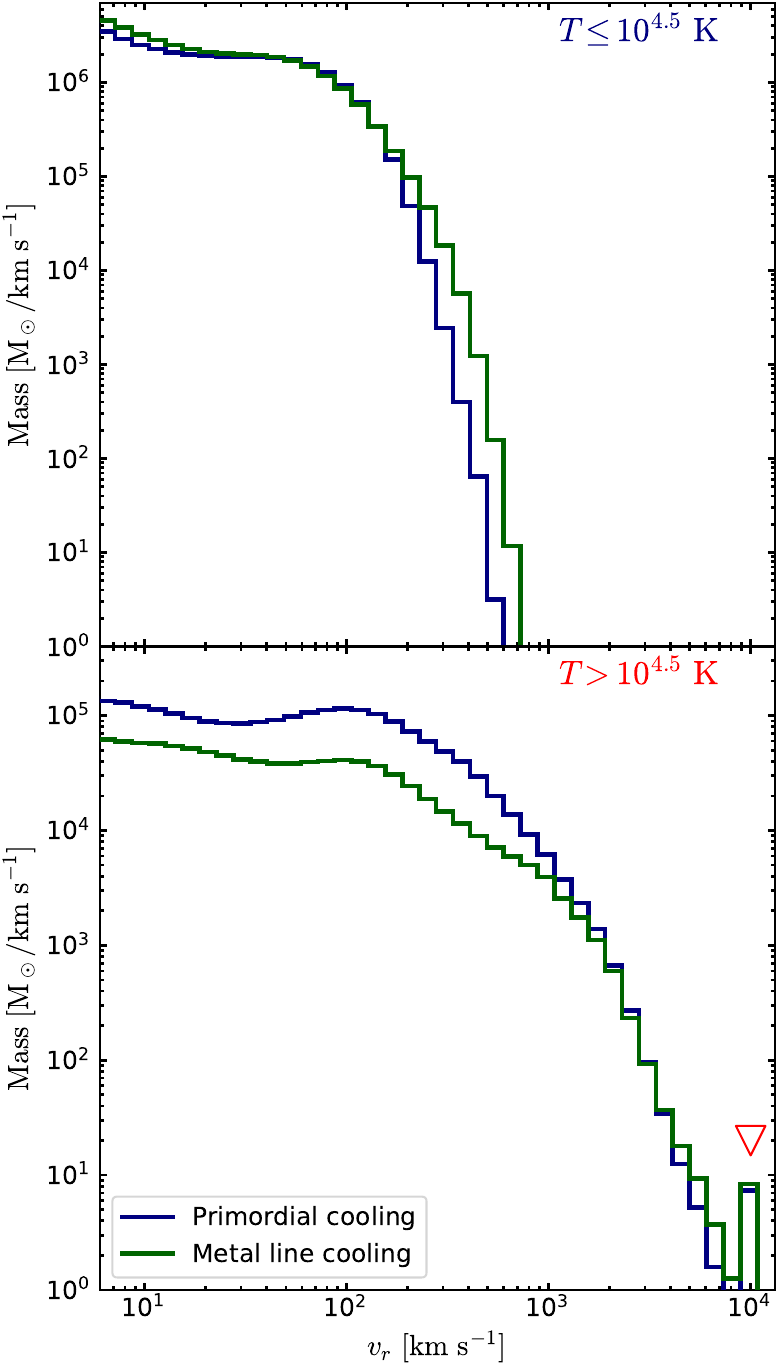}
    \caption{The velocity distribution of the hot (\TEq{>}{4.5}) and cold (\TEq{\leq}{4.5}) outflow components at \tEq{=}{1}. In purple is the fiducial simulation with primordial cooling and in green we show the result when metal line cooling is included. The red triangle marks the wind injection velocity.}
    \label{fig:ap:vr_cooling}
\end{figure}

\begin{figure}
    \centering
    \includegraphics[width=0.47\textwidth]{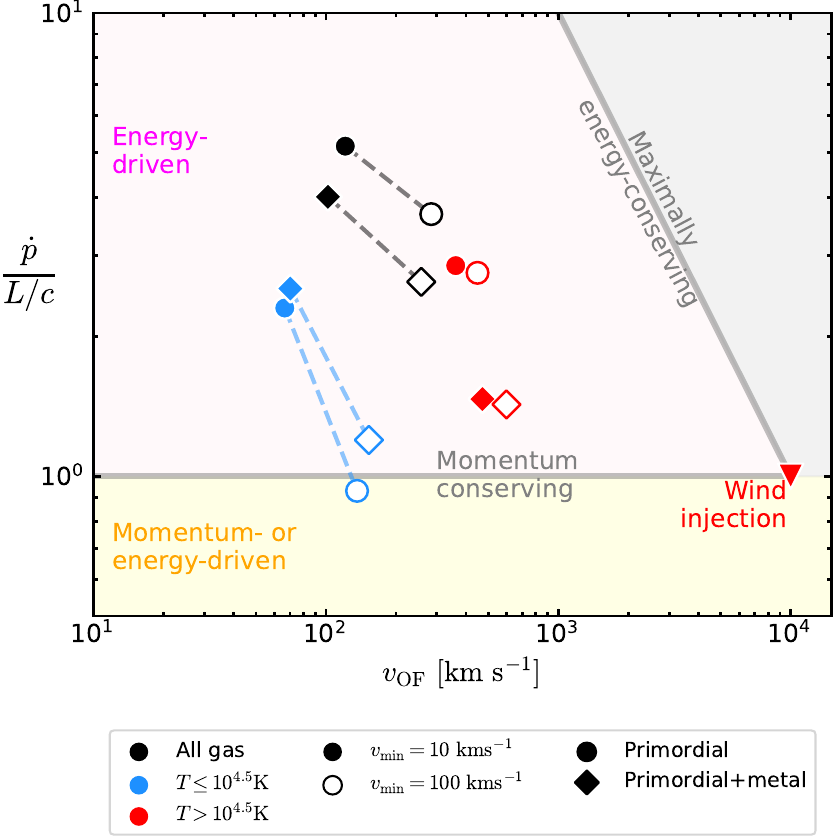}
    \caption{Outflow momentum flux against velocity for the fiducial simulation at \tEq{=}{1}. The circular points are for the simulation with primordial cooling only and the diamonds have primordial and metal line cooling. The colour of the point corresponds to the gas phase. The solid points are for a minimum velocity cut of \vminEq{=}{10} and the empty points are for \vminEq{=}{100}. The velocity is calculated as a mass-weighted mean of all gas cells above the velocity cut. The wind injection velocity and momentum flux is shown as a red triangle. Despite enhanced cooling, the simulation including metal line cooling remains energy-driven (\pEq{\geq}{1}) in all gas phases.}
    \label{fig:ap:energy_driving}
\end{figure}

In \cite{Ward2024} we explored the multiphase outflow resulting from an AGN interacting with a clumpy ISM. In that study we only included primordial cooling, but in this work we also explored including models for metal line cooling. In Figure \ref{fig:ap:vr_cooling} we show the effect this has on the velocity distribution of the hot (\TEq{>}{4.5}) and cold (\TEq{\leq}{4.5}) outflows at \tEq{=}{1} for the fiducial luminosity of \LAGNEq{=}{45}. The increase in cooling efficiency results in a slightly higher outflow mass in the cold phase, and a higher maximum speed for the cooler gas (up to around \vrEq{\simeq}{700}). Likewise, the hot phase loses some mass to the cooler phase for gas moving at \vrEq{\lesssim}{1000}. However, the qualitative conclusion of \cite{Ward2024} of a two-phase outflow with the components moving at different velocities still holds.

In Figure \ref{fig:ap:energy_driving}, we replicate Figure 10 from \cite{Ward2024}, investigating the energy driving of the wind. This is a common plane used by observers \citep[e.g.,][]{Bischetti2019,Bischetti2024,Marasco2020,Longinotti2023} to infer whether the outflow is seen as momentum- or energy-driven. We show the region where \pEq{>}{1} (interpreted as energy-drive) in pink, \pEq{<}{1} in yellow, and the energy- and momentum-conserving lines in grey. The results from the fiducial simulation with only primordial cooling included are marked by circles and the simulations with primordial and metal line cooling are shown by diamonds. The colour of the point refers to the gas phase: black points show all gas, and blue/red show the cold/hot gas phases, split around \TEq{=}{4.5}. As discussed in \cite{Ward2024}, the minimum velocity cut assumed affects the measured outflow energetics significantly, so we show two cuts: \vminEq{=}{10} (a `theoretical' cut; solid points) and \vminEq{=}{100} (a more `observational' cut; hollow points). We can see that, even with enhanced cooling from the metal lines, the outflow is still in the energy-driven regime ($\dot{p}/(L/c)>1$) for all gas phases, with a total momentum flux (assuming \vminEq{=}{10}) of \pEq{=}{4}, compared to a value of \pEq{=}{5} for the primordial cooling only simulation.

\section{Parameter variation} \label{sec:ap:params}

\begin{figure}
    \centering
    \includegraphics[width=0.42\textwidth]{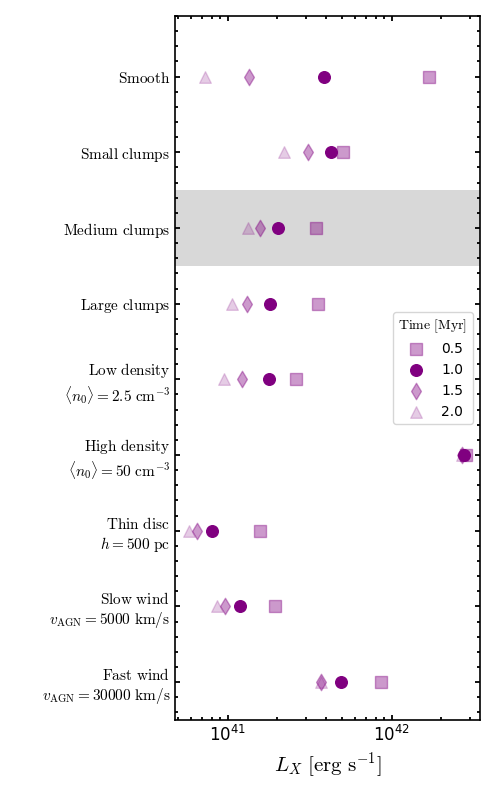}
    \caption{Sensitivity of the total Bremsstrahlung emission produced to various galaxy and AGN properties. There is only minor sensitivity to the initial clump sizes of the ISM, and moderate sensitivity to the density of the ambient medium and to the AGN wind injection velocity.}
    \label{fig:ap:params}
\end{figure}

In Figure \ref{fig:ap:params} we show the time evolution for the Bremsstrahlung X-ray emission for a wider range of model parameters at the fiducial luminosity of \LAGNEq{=}{45} (see Table \ref{table:runs}). As shown in Figure \ref{fig:time_param}, the X-ray emission reaches a peak at around \tEq{\simeq}{0.5} and then slowly decays, and we see a similar trend with for all the runs here. Increasing the size of the initial clumps, or reducing the initial ISM density compared to the fiducial run results in only a small decrease in the total X-ray luminosity produces. Halving the height of the disc decreases the emission by a factor of 2-3, demonstrating that the X-ray emission is probing the total volume of gas the wind has mixed with. Increasing the density of the ambient ISM by a factor of 10 ($\langle n_0 \rangle = 50\ \rm{cm^{-3}}$) results in a order of magnitude higher X-ray emission and a much slower time evolution as the wind is more efficiently constrained by the denser surroundings. Finally, the initial velocity of the AGN wind has a mild impact on the resulting emission with the slower wind resulting in a factor of 2 lower X-ray luminosity and the faster wind boosting the emission by a factor of 2.5. This suggests that the efficiency of the wind-ISM mixing increases when the initial outflow is faster and can more efficiently strip gas from the cold clouds and mix with it.

\section{Numerical Convergence}  \label{sec:ap:res}

Simulations of turbulent mixing and cloud ablation are particularly sensitive to numerical resolution. As turbulence develops, the cloud-wind interface increases, amplifying the effective surface area for mixing \citep{Lancaster2021,Tan2024}. However, this fine structure is often under-resolved, causing part of the energy diffusion between the gas phases to be numerical rather than physical \citep{Lancaster2024}.As we showed in \cite{Ward2024}, increasing the resolution decreased the size of the outflowing clouds; however, we found the global outflow properties to be reasonably well-converged at our fiducial resolution of \MtargetEq{=}{100} (in broad agreement with e.g., \citealt{Gronke2020,BandaBarragan2021,Abruzzo2024}.)

\begin{figure}
    \centering
    \includegraphics[width=0.45\textwidth]{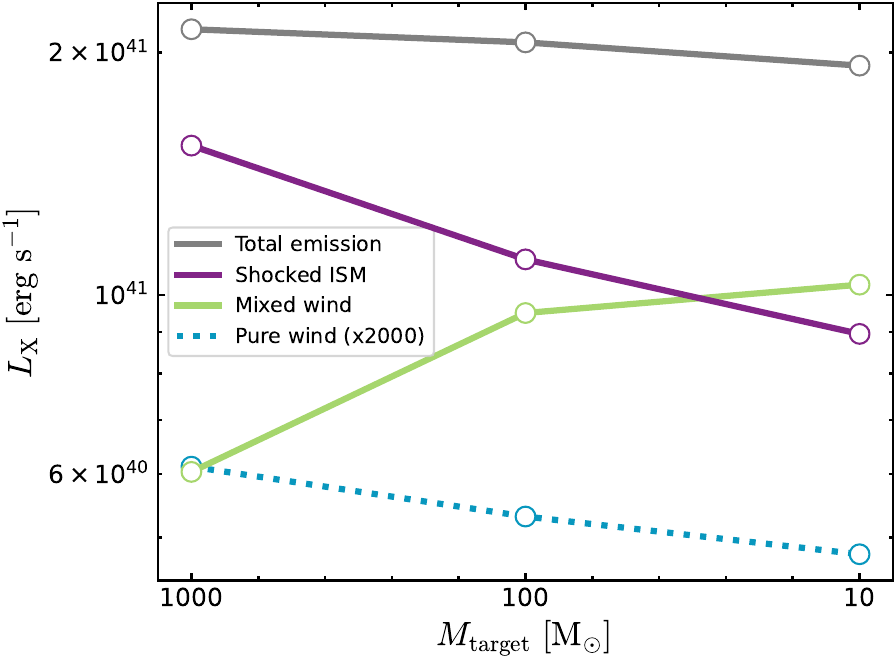}
    \caption{The effect of numerical resolution on the Bremsstrahlung emission predicted for the different wind phases (quantified by the wind tracer density, $\mathcal{P}$)  at \tEq{=}{1}. In purple we show the un-mixed, shocked ISM ($\mathcal{P}<10^{-3}$); in green, the mixed wind-ISM phase ($10^{-3}<\mathcal{P}<0.5$) and in blue we show the pure wind ($\mathcal{P}>0.5$), scaled up by a factor of 2000 for visual comparison.}
    \label{fig:ap:resolution}
\end{figure}

To investigate the convergence of the X-ray emission results presented in this paper, Figure \ref{fig:ap:resolution} shows the Bremsstrahlung emission of our simulations at \tEq{=}{1} for our fiducial setup (\LbolEq{=}{45}; \kminEq{=}{12}; \vAGNEq{=}{10000}). We test three target mass resolutions of 10, 100 (fiducial) and 1000 $\rm{M_\odot}$.
The black line shows the total Bremsstrahlung luminosity which declines only slightly across the three resolutions tested. However, the contribution from the shocked ISM (purple) and pure wind (blue, scaled up by 2000 for plotting clarity) decrease as the resolution improves, whereas the mixed phase (green) increases. This may be due to the increased surface area available for mixing in higher resolution runs; both from the smaller size of the outflowing clouds \citep{Ward2024}, and the better numerical resolution of the turbulent mixing. This interpretation is further supported by our result that the ISM setup with smaller initial clouds also displays an increase in the X-ray emission from the mixed phases compared to the fiducial cloud size. Overall, however, our results are reasonably well-converged between the fiducial ($100\ \rm{M_\odot}$) and high-resolution ($10\ \rm{M_\odot}$) runs for the global X-ray predictions presented in this paper regarding the total Bremsstrahlung emission. Furthermore, the increase in the emission from the mixed phase at finer mass resolutions strengthens our conclusion that this phase is dominant for producing collisionally-excited X-ray emission.

Running the high resolution (\MtargetEq{=}{10}) simulations is computationally expensive, limiting the exploration possible at these resolutions. However, techniques developed by Almeida et al. (sub. MNRAS) for an enhanced refinement scheme to increase resolution in the cold clouds could allow these mixing and ablation effects to be explored at higher resolution, while reducing the overall cost of the simulation.

\bsp	
\label{lastpage}
\end{document}